\documentclass[twocolumn]{aa}  
\usepackage{graphicx}
\usepackage{txfonts}
\usepackage{amsmath,amsfonts,amssymb}
\usepackage{fixltx2e}
\usepackage{color}
\usepackage{natbib}
\bibpunct{(}{)}{;}{a}{}{,}
\usepackage[breaklinks, colorlinks, citecolor=blue]{hyperref}

\usepackage{colortbl}
\definecolor{lightgray}{rgb}{0.83, 0.83, 0.83}

\newcommand{\Planck}{{\em Planck}}

\newcommand{\Herschel}{{\em Herschel}}

\newcommand{\estMd}{\widehat{M}_{\rm dust}}
\newcommand{\estMfive}{\widehat{M}_{500}}

\newcommand{\estLs}{\widehat{L}_{\rm synch}}

\newcommand{\nui}{\nu_i}

\begin{document}

   \title{Quasar host environments: The view from \Planck}

   \author{Lo\"ic Verdier\inst{1}
          \and
          Jean-Baptiste Melin\inst{1}
          \and
          James G. Bartlett\inst{2}\fnmsep\inst{3}
          \and %{\bf (order TBD)}
          Christophe Magneville\inst{1}
          \and
          Nathalie Palanque-Delabrouille\inst{1}
          \and
          Christophe Y\`eche\inst{1}
         }

   \institute{DSM/Irfu/SPP, CEA-Saclay, F-91191 Gif-sur-Yvette Cedex, France\\
        \email{loic.verdier@cea.fr, jean-baptiste.melin@cea.fr}
                      \and
            APC, AstroParticule et Cosmologie, Universit\'e Paris Diderot, CNRS/IN2P3, CEA/lrfu, Observatoire de Paris, Sorbonne Paris Cit\'e, 10, rue Alice Domon et L\'eonie Duquet, 75205 Paris Cedex 13, France\\
            \email{bartlett@apc.univ-paris7.fr}
            \and
            Jet Propulsion Laboratory, California Institute of Technology, 4800 Oak Grove Drive, Pasadena, California, U.S.A.
            }

   \date{Received 23 September 2015 / Accepted 20 January 2016}

% \abstract{}{}{}{}{} 
% 5 {} token are mandatory
 
  \abstract{We measure the far-infrared emission of the general quasar (QSO) population using \Planck\ observations of the Baryon Oscillation Spectroscopic Survey QSO sample.  By applying multi-component matched multi-filters to the seven highest \Planck\ frequencies, we extract the amplitudes of dust, synchrotron, and thermal Sunyaev-Zeldovich (SZ) signals for nearly 300 000 QSOs over the redshift range $0.1<z<5$.  We bin these individual low signal-to-noise measurements to obtain the mean emission properties of the QSO population as a function of redshift.  The emission is dominated by dust at all redshifts, with a peak at $z \sim 2$, the same location as the peak in the general cosmic star formation rate.  Restricting analysis to radio-loud QSOs, we find synchrotron emission with a monochromatic luminosity at 100\,GHz (rest-frame) rising from $\overline{L_{\rm synch}}=0$ to $0.2 \, {\rm L_\odot} {\rm Hz}^{-1}$ between $z=0$ and 3. The radio-quiet subsample does not show any synchrotron emission, but we detect thermal SZ between $z=2.5$ and 4; no significant SZ emission is seen at lower redshifts. Depending on the supposed mass for the halos hosting the QSOs, this may or may not leave room for heating of the halo gas by feedback from the QSO.  
%We used the BOSS (Baryon Oscillation Spectroscopic Survey) quasars as tracers of halos between redshift $z \sim 0$ and 5, and studied their millimeter emission in the \Planck\ maps. The total emission is composed of a mixture of dust, synchrotron, and thermal Sunyaev-Zel'dovich (SZ) effect that we extracted using multi-component Matched Multi-Filters. . Assuming that the emission arise from the thermalized gas in the halo and that the local SZ-mass scaling relation remains valid in this redshift range, we constrain the average halo mass $\overline{M_{500}} = (2.5 \pm 0.3) \times 10^{13} M_\odot$. It is nevertheless possible that part of the detected SZ signal originate from non-thermalized gas ejected by the quasar in the halo. The above mass should then be considered as an upper limit on the halo mass hosting quasars in $2.5<z<4$.
}

   %{Here is the context}
   %{Here are the aims}
   %{Here are the methods}
   %{Here are the results}
   %{Here are the conclusions}
  
   %The baryonic hot gas, the main baryonic component of the haloes has been detected 

   \keywords{ cosmology: observations
   		- large-scale structure of Universe
		- quasars: general
		- galaxies: clusters: general
		- methods: data analysis
		- methods: statistical
               }

   \maketitle
%
%________________________________________________________________

\section{Introduction}

Quasars (or quasi stellar objects, QSOs) occupy a special place in large-scale structure and galaxy evolution~\citep{kormendy1995}.  
They are among the most luminous extragalactic sources, and, as such, have become  the focus of many cosmological surveys such as the 2dF Quasar Redshift Survey~\cite[2QZ;][]{Croom2001} or   the successive iterations of the Sloan Digital Sky Survey~\citep[SDSS~I-IV;][]{york2000, Eisenstein2011}, where they provide a unique insight into the formation of structure on large scales and at high redshift~\citep{White2012}. 
Quasars are likely powered \citep{Salpeter1964,Lynden1969} by accretion of nearby matter onto supermassive black holes \citep[SMBH,][]{rees1984}, whose evolution appears closely linked to the general cosmic star formation rate \citep{madau2014}, in particular in galaxies that contain a massive bulge (and therefore a massive central black hole) and a gas reservoir~\citep{Nandra2007, Silverman2008}.  The link is a clue to galaxy formation, its significance emphasized by the observed relation between the SMBH mass and galaxy properties \citep{ferrarese2000, gebhardt2000}.  In addition, galaxy formation models must evoke strong feedback from AGN to explain the observed properties of massive galaxies and to avoid the overcooling catastrophe \citep{croton2006, bower2006, mcnamara2007, somerville2008, fabian2012, blanchard1992}

Quasar environments give us a look at these powerful engines, and millimeter/submillimeter observations can offer a particularly revealing view of their effects: at these frequencies it is possible to study the cooler dust emission associated with star formation in the host, look for synchrotron emission from energetic particles, and, perhaps most pertinently, to measure energy feedback through observation of the thermal Sunyaev-Zeldovich \citep[tSZ,][]{sz1970,sz1972} effect, a direct probe of the thermal energy contained in surrounding gas.  Samples observed in this waveband with large ground-based facilities typically consist of, at most, several tens of objects \citep{omont2001, isaak2002, omont2003, priddey2003, 2006ApJ...642..694B, wang2007, omont2013}.  Operating at these same frequencies, cosmic microwave background (CMB) surveys cover large sky areas and present the opportunity of studying much larger samples, albeit with much less detail of individual objects.  We can, however, determine the average properties of large representative populations, a valuable compliment to the more detailed examinations. 

The Wilkinson Microwave Anisotropy Probe \citep[WMAP,][]{bennett2003} and \Planck\ missions \citep{tauber2010, planck2011-p1} are ideal for this purpose because of their all-sky coverage, including the entire Sloan Digital Sky Survey \citep[SDSS,][]{york2000} area.  Cross-correlating WMAP data with photometric QSOs from SDSS Data Release 3 (DR3), \citet{chatt2010} finds evidence at 2.5$\sigma$ for the tSZ effect from QSO environments.  \citet{ruan2015} improved the significance using a publicly available tSZ map \citep{hill2014} constructed from the \Planck\ mission dataset and 26 686 spectroscopic QSOs from SDSS DR7.  They concluded that the total thermal energy feedback into surrounding gas was significantly larger than expected according to galaxy formation models, which is consistent with the findings of \citet{chatt2010}.  \Planck's wide frequency coverage, from 30\,GHz to 857\,GHz, is a distinct advantage for disentangling the different sources of emission from the QSO environment, and was implicitly exploited by \citet{ruan2015} when using the tSZ map published by \citet{hill2014}.  

Although not all-sky, the Atacama Cosmology Telescope (ACT) surveyed several hundred square degrees near the celestial equator, overlapping AGN and QSO samples in the North. \cite{2014MNRAS.445..460G} stacked millimeter maps from ACT over this area of radio-loud AGN seen in the FIRST and NVSS radio surveys to find a tSZ signal at $5\sigma$ significance.  More recently, and in parallel to our present study, \cite{2015arXiv151005656C} combined ACT and {\em Herschel} data of radio-quiet quasars in the SDSS DR7 and DR10 samples, detecting the tSZ signal at $3-4\sigma$ significance.

In this paper, we use the full-mission \Planck\ frequency maps to examine the emission properties of the SDSS-III Baryon Acoustic Oscillation \citep[BOSS,][]{dawson2013} DR12 QSO sample.  We extract not only the tSZ signal from the population, but also dust and synchrotron emission by stacking measurements made with a set of matched filters directly applied to the seven highest  frequency \Planck\ channels.  The joint extraction enables us to study not only the gas thermal environment, but star formation conditions and the production of energetic particles.  Moreover, it improves analysis robustness by giving information on correlations between the observed signals, compared to the use of a tSZ map.  \citet{ruan2015} were in fact faced with the difficulty of correcting the SZ map for contamination from dust emission.  

We extend our multi-matched filter \citep[MMF,][see also \cite{2002MNRAS.336.1057H}]{2006A&A...459..341M} formalism to simultaneously extract two or three emission components, following development started in \citet{lbg2013}.  Stacking the various signals to study their evolution with redshift, we reconstruct for the first time a picture of the evolution of the dust signal, the synchrotron emission, and the tSZ effect between $z \sim 0$ and $z \sim 5$ for the general QSO population.

In Sect.~\ref{sec:data}, we describe the BOSS and \Planck\ datasets. We detail the analysis methods and tools in Sect.~\ref{sec:analysis}. In Sect.~\ref{sec:simu}, we apply them on simulations. Results on \Planck\ data are given in Sect.~\ref{sec:results}. We discuss our results in Sect.~\ref{sec:discussion} and conclude in Sect.~\ref{sec:conclusion}.  Throughout, we use the spatially flat base $\Lambda$CDM cosmology from \citet{planck2015-xiii}: $H_0=67.27 = h 100$\,km/s/Mpc, $\Omega_{\rm m}=0.3156$, $\Omega_{\rm b} h^2=0.02225$ and $\sigma_8=0.831$.

%__________________________________________________________________

\section{Data}
\label{sec:data}

\subsection{BOSS quasars}
\label{BOSS quasars}

%\jbm{I suggest that this section is re-written/improved by the BOSS team in Saclay. Nathalie or Christophe ?}
%Quasi-Stellar Objects (QSO or quasars) are very bright optical-infrared sources with typical luminosities ten times to hundred thousands times the luminosity of the Milky Way. \jbm{in which spectral domain: optical, radio ?}\jbm{references needed} They are galaxies hosting a supermassive black hole fueled by the accretion of nearby matter. QSO seem to live in denser environment than standard galaxies and could be possible tracers for the presence of galaxy clusters \jbm{references needed}.

One of the major goal of the Sloan Digital Sky Survey-III (SDSS-III) Baryon Oscillation Spectroscopic Survey (BOSS) is the production of a QSO catalogue to detect the BAO scale in the \mbox{Lyman-$\alpha$} forests at redshift~\mbox{z$\sim 2.5$}. A first detection was made on the DR9 QSO catalogue~\citep{paris2012} by~\cite{busca2013}, confirmed latter by~\cite{delubac2015}.  In this study, we use the recently published DR12 QSO catalogue\footnote{\url{http://www.sdss.org/dr12/algorithms/boss-dr12-quasar-catalog/}}~\citep{alam2015}.

The sources are detected and selected with the CCD imaging Camera installed in the Sloan Foundation 2.5 m Telescope at Apache Point observatory, New Mexico. The spectra of each source is measured with the BOSS spectrograph which covers wavelength between 3600~$\AA$ and 10 000~$\AA$. Spectra measurement leads to the computation of the spectroscopic redshift and the checking of the nature of the source.\\

\begin{figure}
\centering
\includegraphics[width=.5\textwidth]{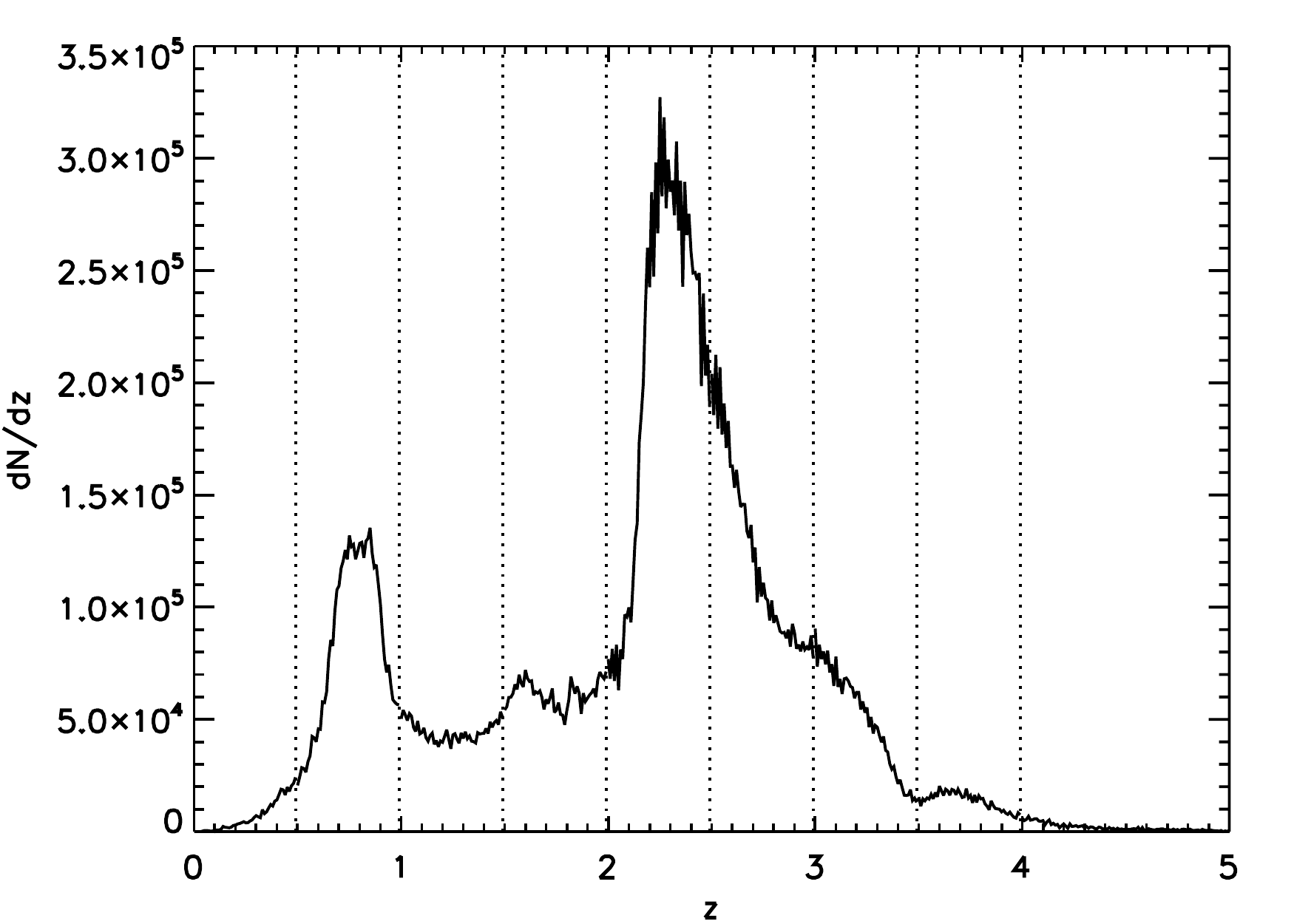}
\caption{Redshift distribution of the BOSS DR12 QSO sample. The distribution presents two peaks, at $z\sim 0.8$ and $z\sim 2.3$, due to the QSO target selection process (see text). The regular binning used throughout this paper ($\Delta z=0.5$) is shown as the vertical dotted lines.}
\label{fig:nz_qso}
\end{figure}

The DR12 QSO catalogue contains 297 301 objects. We remove 5880 QSO falling outside the \Planck\ 65$\%$ mask, in a region strongly contaminated by the Milky Way dust. Finally, we also reject 256 QSO because they are at low redshift $z<0.1$ so may be partially resolved by \Planck\,, because they belong to the high redshift  (z>5) population or because they have a bad estimate of the magnitude in g band ($g<0$) . We thus use a catalogue of 291 165 QSO in the redshift range $0.1<z<5$. The distribution of the QSO in term of redshift is not flat and reach two notable maxima at z=0.8 and z=2.3 (Fig.~\ref{fig:nz_qso}). 
%\jbm{We'll check later if this condition is really necessary. Do we want to present plots as a function of magnitude in g band or not ?}\lv{Add plots is not necessary or only the plot for the dust. But I think it's worth mentioning we have checked the evolution of our different masses across g and we don't have significant evolution. We have only a probably "non-intrinsic" decrement of our dust flux (the most notable) and synchrotron flux} 

The target selection was developed to select quasars with an observable Ly-$\alpha$ forest (i.e quasars with z>2.2). However, degeneracies in the color-redshift relation of quasars led to the selection of low-z quasars in BOSS (Fig. ~\ref{fig:nz_qso}). The quasars at $z \sim 0.8$ have MgII $\lambda$2800 line at the same wavelength as  Ly-$\alpha$ at redshift $z \sim 3.1$, giving these objects similar broadband colors, while the large number of objects at $z \sim 1.6$ is due to the confusion between $\lambda$1549 C-IV line and Ly-$\alpha$ at $z \approx 2.3$  ~\citep{ross2013}. In contrast, the selection based on the intrinsic  variability of quasars gives a more uniform distribution in redshift ~\citep{palanque2011}.
%\jbm{Nathalie, Christophe, we should discuss here why there are two maxima.}

\subsection{\Planck\ maps}
The \Planck\ satellite was launched in May 2009~\citep{tauber2010}. After traveling to the Earth-Sun Lagrange point L2, it scanned the entire sky continuously from August 2009 to October 2013 in nine frequency bands ranging from 30 to 857\,GHz. The primary goal of the mission was to study the primary CMB anisotropies, but its large frequency coverage also enables unique Galactic and extra-galactic  astrophysical studies.  In particular, the High Frequency Instrument \citep[HFI,][]{lamarre2010, planck2011-1.5}, covering bands from 100 to 857\,GHz, is ideally suited for Sunyaev-Zeldovich science. Moreover, the three highest frequencies (353 to 857\,GHz) of the instrument pick up Galactic and extra-galactic dust emission.  The Low Frequency Instrument \citep[LFI,][]{bersanelli2010, planck2011-1.4}, with channels at 30, 44, 70\,GHz, is sensitive to synchrotron, free-free, and spinning dust emission.

We use the seven highest frequency (70, 100, 143, 217, 353, 545, and 857\,GHz) full-sky maps from the 2015 data release~\citep{tauber2015,tristram2015,elina2015}.  These bands are the best suited to study the dust, gas, and synchrotron emission from high redshift QSO environments.  The substantially larger beams of the remaining 30 and 40\,GHz channels make them less well adapted for studying what are essentially point sources.  We divide each \Planck\ HEALPix\footnote{\url{http://healpix.sourceforge.net}} all-sky map into 504 overlapping flat tangential patches of $10\times10$ degrees with a $1.72~{\rm arcmin}$ pixel scale in order to apply our extraction algorithms described in Sec.~\ref{sec:mf}. 

\section{Analysis}
\label{sec:analysis}
Determining the physical properties of the QSO environment from the \Planck\ maps faces two major challenges. The first is the faint QSO flux in the \Planck\ maps, far below the noise level; individual detection is not possible. We thus have to use a statistical approach similar to the one developed in~\cite{melin2011} and subsequently in~\cite{pxcc2011,maxbcg2011,lbg2013}. Details of the approach are given in Sect.~\ref{sec:avg}. 

The second challenge arises from the superposition of different emission sources from the QSO in the \Planck\ beams:  emission from  dust located inside the host galaxy or possibly at larger scale in the host halo of the QSO; synchrotron emission from the host galaxy or from relativistic outflows of the central AGN; and the tSZ effect due to the hot gas surrounding the QSO and within the host halo.  \Planck\ does not have the spatial resolution to separate these components, but its good spectral coverage enables us to disentangle these sources of emission.

We separate the signals using multi-component matched multi-filters (MMF), an extension of the approach used in~\cite{lbg2013}. The detailed description is given in Sec.~\ref{sec:mmf}.  We will see that the QSO signal in \Planck\ is dominated by dust emission, but that we also detect both synchrotron and tSZ signals.  In practice, we proceed as follows:
\begin{enumerate}
\item Assume that the QSO signal is a mixture of one, two or three components;
\item Apply the adapted multi-component MMF at each QSO position to obtain the amplitude of each component;
\item Bin average (as a function of redshift or magnitude) these amplitudes over the QSO sample;
\item Evaluate the ability of our model to describe the data using a $\chi^2$-test (see Sect.~\ref{sec:chi2-test for the hypothesis}).
\end{enumerate}

Our notation in the subsequent sections is as follows.  We use the letter $p$ as index to denote individual QSOs in the catalogue and the indices $i$ and $j$ to denote observation frequencies.  The Greek indices $\lambda$ and $\mu$ specify the nature of the emission components, e.g., $\lambda$ and $\mu$ have as possible values {\em dust, synch}, and {\em tSZ} for dust, synchrotron, and  tSZ signals, respectively.

\subsection{Model of the QSO emission}
\label{sec:qso_model}
We model the QSO emission as a sum of dust,  synchrotron, and tSZ signals.  
%Thereafter, we suppose one or two of this components to be null.
The seven-frequency column vector map, $\vec{m}(\vec{x})$, at position $\vec{x}$ on the sky is written as
\begin{equation}
%\begin{split} 
\label{eq:datamodel}
   \vec{m}(\vec{x}) = \sum_{\lambda}a_{\lambda} \, \vec{t}_{\lambda}(z,\vec{x}) + \vec{n}(\vec{x}),
%\end{split}
\end{equation}
where $a_{\lambda} $ is the amplitude of the $\lambda$ component, $\vec{t}_{\lambda}(z,\vec{x})$ is the associated, normalized emission vector, $z$ the redshift of the QSO, and $\vec{n}(\vec{x})$ is the astrophysical and instrumental noise.  We center the QSO in the map to simplify the expressions. The $i^{th}$ component of each emission vector is the signal profile, $\tau_{\lambda}$ (normalized to unity at the origin), convolved by the \Planck\ beam, $b_i$ (normalized to unity at the origin), at frequency $\nui$, and then scaled by the expected frequency dependance $\left(S_{\lambda}\right)_i(z)$:
\begin{eqnarray}
  \left(t_{\lambda} \right)_i(z,\vec{x}) =  \left(S_{\lambda}\right)_i(z)\, [b_i\ast \tau_{ \lambda}](\vec{x}).
\end{eqnarray}

We describe the dust emission frequency dependence by a modified blackbody, 
%($T_{\rm d},\beta_{\rm d}$),
%~\citep[see e.g.,][]{2006ApJ...642..694B}, 
the synchrotron emission by a power law, and we use the non-relativistic calculation for the tSZ  \citep[e.g.,][]{1999PhR...310...97B,2002ARA&A..40..643C}:
\begin{eqnarray}
   \left(S_{\rm d}\right)_i(z) = \left ({\nui \over 857 {\rm GHz}} \right )^{\beta_{\rm d}} {B_{\nui}[T_{\rm d}/(1+z)] \over B_{857 {\rm GHz}} \, [T_{\rm d}/(1+z)]}, \\
  \left(S_{\rm s}\right)_i= \left ( {100 {\rm GHz} \over \nu_i } \right )^{\alpha_{\rm s}},\\
  \left(S_{\rm sz}\right)_i= B_{\nui}(T_{\rm cmb}){x_i e^{x_i} \over e^{x_i}-1} \left [ {x_i \over \tanh(x/2)} - 4 \right],
\end{eqnarray}
where $x_i= h \nui/k T_{\rm cmb}$ with $T_{\rm cmb}$ the temperature of the CMB today, $B_\nu(T)$ represents the Planck spectrum, $T_{\rm d}$ and $\beta_{\rm d}$ are the dust temperature and emissivity index, and $\alpha_{\rm s}=0.7$ for the synchrotron index. It is the typical value for radio galaxies \citep{1992ARA&A..30..575C,1997iagn.book.....P}. We note that the first two expressions are unitless, so that $a_{\rm dust}$ and $a_{\rm synch}$ carry units of brightness; on the other hand, $\vec{S}_{\rm tSZ}$ carries brightness units while $a_{\rm tSZ}$ is unitless and corresponds to the central Compton-$y$ parameter.

%$\Tdu(\vec{x})$ and $\Tsy(\vec{x})$ are the profiles for dust and synchrotron signals. 
We adopt the same spatial profile for all signal templates, $\tau_\lambda = \tau(x/\theta_{\rm f})$, where $\tau$ is taken  
%that they both follow the gas spatial distribution in the halo, modeled with $\Tth(\vec{x})=\Tth(x/\thetat)$ 
from~\cite{2010A&A...517A..92A} based on the observed cluster halos gas pressure profile.  We fix $\theta_{\rm f}= 0.27$\,arcmin, the value corresponding to a halo of mass $M_{500}=10^{13}M_{\odot}$ at $z=2$.
%assuming the \cite{2010A&A...517A..92A} pressure profile is still valid at this redshift. 
This value is smaller than the smallest \Planck\ beam (${\rm fwhm}=4.2\,{\rm arcmin}$ at 857~GHz): the bulk of the QSO remain unresolved by \Planck. We study the sensitivity of our results to this fixed size in Sect.~\ref{sec:syssize}.  The profiles are cut at $\theta > 5\theta_{500}$.

In the following we therefore denote this universal QSO profile by $\tau_{\rm qso}$.   We are then able to define the QSO brightness, $\vec{f}$, as the sum over the different components as
\begin{equation}
f_{\rm i}=\sum_{\lambda}{a_{\lambda} \, \left(S_\lambda \right)_i}(z),
\end{equation}
and rewrite the map at frequency $i$ as
\begin{equation}
m_i(\vec{x}) = f_i \, \left(t_{\rm qso}\right)_i(\vec{x}) + n_i(\vec{x}),
\end{equation}
where $\left(t_{\rm qso} \right)_i (\vec{x}) \equiv [b_i \ast \tau_{\rm qso}](\vec{x})$.

\subsection{Matched filters}
\label{sec:mf}

\subsubsection{Single frequency matched filter}
We use single frequency matched filters to extract individual QSO signals (at $S/N \ll 1$) at each \Planck\ frequency. The individual signals are necessary to compute the $\chi^2$ statistic described in Sect.~\ref{sec:chi2-test for the hypothesis}.  We also average them to obtain the mean QSO spectrum in Sect.~\ref{sec:results}. 

The single frequency matched filter is a linear filter designed to return an unbiased signal estimate with minimal error~\citep{ht1996, 2001ApJ...552..484S}. No assumption is made concerning the frequency dependance of the signal. The estimated value, $\hat{f}_{\rm i}$, of the true signal, $f_{\rm i}$, is
\begin{equation}
 \hat{f}_{\rm i}= \int d^2x \; \Psi_{\rm i}^t(\vec{x}) \, m_{\rm i}(\vec{x}),
\end{equation}
where $\Psi_{\rm i}(\vec{x})$ is the matched filter for frequency $\nui$.  The filter $\Psi_{\rm i}$ is expressed in Fourier space as
\begin{equation}
\Psi_{\rm i}(\vec{k}) =  \sigma^2[ \hat{f}_{\rm i}] \frac{\left(t_{\rm qso}^{*}\right)_i(\vec{k}) }{P_{ii}(\vec{k})},
\end{equation}
with the error on the estimated signal 
\begin{equation}
\begin{split}
\label{eq:sigt_flux}
    \sigma[\hat{f}_{\rm i} ]      & \equiv  \sqrt{ \langle{ \hat{f}_{\rm i}^2 \rangle} - \langle{ \hat{f}_{\rm i} \rangle}^2} \\
    & =   \left[\int d^2k\; 
     \frac{ \left| \left(t_{\rm qso}\right)_i(\vec{k}) \right|^2}{P_{ii}(\vec{k})}\right]^{-1/2},
\end{split}     
\end{equation}
where $P_{ii}(\vec{k})$ is the noise power spectrum of the $i^{th}$ \Planck\ map, $m_i(\vec{k})$. Because the signal-to-noise of an individual QSO signal is smaller than unity, we estimate $P_{ii}(\vec{k})$ directly as the power spectrum of the map.

\subsubsection{Matched Multi-Filters (MMF)}
\label{sec:mmf}
The matched multi-filters (MMF) were first introduced by~\cite{2002MNRAS.336.1057H} for SZ detection.  They were further developed by~\citet{2006A&A...459..341M} and extensively used to construct the successive \Planck\ SZ cluster catalogs~\citep{esz2011,psz1,psz2}. The MMF target the extraction of a single component across a set of maps, assuming a known frequency dependance and a spatial distribution.  The filtered map returns a unbiased estimate, $\hat{a}_{\lambda}$, of $a_{\lambda} $ with minimal variance:
\begin{equation}
\hat{a}_{\lambda} = \int d^2x \; \vec{{\Psi}}_{\lambda}^t(\vec{x}) \cdot \vec{m}(\vec{x}).
\end{equation}
%As mentioned previously, the MMF were introduced  to extract the tSZ signal, so for $\lambda = \rm{tSZ}$.  In this study, dust emission dominates and we use the MMF assuming our signal to come from the dust only ($\lambda = \rm{\rm dust}$). 
Similarly to the single frequency case, the MMF for the $\lambda$ emission, $\vec{{\Psi}}_{\lambda}(\vec{x}), $ is expressed in Fourier space as
\begin{equation}
\vec{{\Psi}}_{\lambda}(\vec{k}) = \sigma^2[\hat{a}_{\lambda}] \, \vec{P}^{-1}(\vec{k})\cdot \vec{t_{\lambda}^{*}}(z,\vec{k}).
\end{equation}
The error on the amplitude is now given as
\begin{equation}
\begin{split}
\label{eq:sigt}
\sigma[\hat{a}_{\lambda}]        & \equiv \sqrt{\langle{ \hat{a}_{\lambda}^2 \rangle} - \langle{ \hat{a}_{\lambda} \rangle}^2} \\
    &= \left[\int d^2k\; 
     \vec{t_{\lambda}^{*}}^{t}(z,\vec{k})\cdot \vec{P}^{-1}(\vec{k}) \cdot
     \vec{t_{\lambda}}(z,\vec{k})\right]^{-1/2},
\end{split}     
\end{equation}
with $\vec{P}(\vec{k})$ being the inter-band cross-power spectrum matrix with contributions from (non-${\lambda}$) sky signal and instrumental noise.  It is the effective noise matrix for the MMF and, as for the single frequency power spectrum, can be estimated directly on the data, since the $\lambda$ signal (dust, synchrotron or tSZ) is small compared to other astrophysical signals over the sky patch.

Introduced in~\citet{lbg2013} for a mixture of dust and tSZ signals, multi-component filtering deals with different MMF to separate dust and tSZ signals.  Multi-component filtering has also been successfully used to jointly extract the tSZ and kinetic SZ (kSZ) effects in~\citet{kszplanck2014}. For these filters, the recovered amplitude of each component is unbiased with minimal variance if the assumption on the number and type of the components is correct. 

We consider three multi-component filtering schemes: 1) dust+tSZ, 2) dust+synchrotron, and 3) dust+tSZ+synchrotron filters.  The amplitudes $a_{\lambda} $ are estimated using a linear combination of the partial MMFs,
\begin{equation}
\hat{a}_{\lambda} = \sum_{\mu}{ \left[\vec{D}^{-1} \right]_{\lambda, \mu}\, \int d^2x \; \vec{{\Phi}}_{\mu}^t(\vec{x}) \cdot \vec{m}(\vec{x})},
\end{equation}
with the partial MMF defined in Fourier space by
\begin{equation}
\vec{\Phi}_{\mu}(\vec{k}) =  \vec{P}^{-1}(\vec{k})\cdot \vec{t_{\mu}}(z,\vec{k}),
\end{equation}
and the matrix $\vec{D}$ computed as
\begin{equation}
D_{\lambda, \mu}         = \left[\int d^2k\; 
     \vec{t_{\lambda}^{*}}^t(z,\vec{k})\cdot \vec{P}^{-1}(\vec{k}) \cdot
     \vec{t_{\mu}}(z,\vec{k})\right].
\end{equation}

The estimated amplitudes are combined together in a $N$-component vector ($N$ being the number of signal components considered), $\hat{\vec{a}}$. They are correlated with covariance matrix
\begin{equation}
\begin{split}
\label{eq:sigt_cross}
     C[\hat{\vec{a}}]_{\lambda, \mu}  & \equiv  \langle{ \hat{a}_{\lambda} \hat{a}_{\mu} \rangle}  - \langle{ \hat{a}_{\lambda} \rangle}\langle{ \hat{a}_{\mu}\rangle}  \\
    & = \left[\vec{D}^{-1} \right]_{\lambda, \mu}.
\end{split}
\end{equation}

In summary, we use the single frequency matched filter plus the following five filters in our study:
\begin{itemize}
\item Single component MMF for dust;
\item Single component MMF for tSZ;
\item Two-component MMF for dust+tSZ;
\item Two-component MMF for dust+synchrotron;
\item Three-component MMF for dust+tSZ+synchrotron.
\end{itemize}

\subsection{From observed signals to physical QSO properties}
The amplitudes $\hat{a}_{\lambda}$ (determined with profiles cut at $\theta < 5 \theta_{500}$) are converted into spherically integrated quantities within $R_{500}$ by
\begin{eqnarray}
%\widehat{\rm A}_{\lambda} = \hat{\rm a}_{\lambda} \, \int d^2x \; \tau_{\lambda}(\vec{x}) 
\widehat{A}_{\lambda} = \hat{a}_{\lambda} \, %\int_{r<R_{500}} dr \; 4\pi r^2 \tau_{\lambda}(r)
\int_{\theta < 5 \theta_{500}} d\Omega \, \tau_\lambda (\theta) \times C_\lambda(5R_{500} \rightarrow R_{500}),
\end{eqnarray}
where the factor $C_\lambda(5R_{500} \rightarrow R_{500})$ is the conversion from the volume within $5R_{500}$ to $R_{500}$, given the adopted profile.  Similarly, we obtain the source flux density vector, $\widehat{\vec{F}}$, as
\begin{equation}
% \widehat{\vec{F}}=  \hat{\vec{f}}\, \int d^2x \; \tau_{\rm qso}(\vec{x})
\widehat{\vec{F}} = \hat{\vec{f}}\, \, %\int_{r<R_{500}} dr \; 4\pi r^2 \tau_{\rm qso}(r)
\int_{\theta < 5 \theta_{500}} d\Omega \, \tau_{\rm qso} (\theta) \times C_{\rm qso}(5R_{500} \rightarrow R_{500})
\end{equation}
We express the quantities $\widehat{A}_{\rm dust} $, $ \widehat{\vec{F}}$, and $\widehat{A}_{\rm synch} $ in mJy and $\widehat{A}_{\rm tSZ} $ in ${\rm arcmin}^2$.

Following~\cite{2006ApJ...642..694B} we estimate the total dust mass of the QSO from $\widehat{\rm A}_{\rm dust}$ as
\begin{equation}
\estMd=\widehat{A}_{\rm dust}{D_{\rm L}(z)^2 \over (1+z)}\kappa^{-1}\left[857\,{\rm GHz}(1+z)\right]B^{-1}_{857\, {\rm GHz} (1+z)}[T_{\rm d}],
\label{eq:mdust}
\end{equation}
with $\kappa(\nu) = \kappa_0 \left ( {\nu \over 249.8 {\rm GHz}} \right  )^{\beta_d}$, $\kappa_0=0.4 {\rm g}^{-1} {\rm cm^2}$ and $D_{\rm L}(z)$ the luminosity distance.
From the synchrotron flux density at 100\,GHz, we compute the monochromatic synchrotron luminosity according to
\begin{equation}
\estLs=\widehat{A}_{\rm synch} \,{D_{\rm L}(z)^2 \over (1+z)(1+z)^{\alpha_{\rm s}}} \, 
\label{eq:lsynch}
\end{equation}
which is referenced to the rest-frame frequency of 100\,GHz assuming our power law with $\alpha_{\rm s}$.
%Following Eq.~(22) of~\cite{2010A&A...517A..92A}, the total mass of the halo\footnote{We compute $\estMfive^{\alpha}$ because it is proportional to $\widehat{A}_{\rm tSZ}$ which is expected to be Gaussian distributed, allowing for inverse-weighted variance averaging over the full QSO population.} can be estimated by
%\begin{equation}
%\estMfive^{\alpha}= \left[ (1-b) \widehat{M}_{500,{\rm true}} \right ]^{\alpha} = \left [3 \times 10^{14} \left ( {h \over 0.7} \right ) ^{-1} \right ]^{\alpha}  {\widehat{A}_{\rm tSZ}  E(z)^{-2/3} \over A_{\rm x}} M_{\odot}^{\alpha},
%\label{eq:mhalo}
%\end{equation}
%with $\alpha=1.78$ and $A_{\rm x}=2.925 \times 10^{-5} \times 0.6145 \times \left ( {h \over 0.7} \right ) ^{-1} \times D_{\rm A}^{-2}(z) \times \left ( {\pi \over 180} {1 \over 60}  \right ) ^{-2} {\rm arcmin}^2$, $D_{\rm A}(z)$ being the angular diameter distance.  The so-called mass bias parameter, $b$, accounts for any bias in the calibration of this relation, for example from violation of hydrostatic equilibrium and/or X-ray instrument calibration.  \Planck\ SZ cluster counts require high values of the mass bias $(1-b)\sim 0.6-0.7$ for consistency with the \Planck\ base $\Lambda$CDM cosmological parameters determined from the primary CMB anisotropies \citep[see][and references therein]{planck2013-XX, planck2015-XXIV}.

We define the intrinsic Compton parameter, $\widehat{Y}_{500}$, as the quantity $\tilde{Y}_{500}$~from~\cite{lbg2013},
\begin{equation}
\widehat{Y}_{500} = \widehat{A}_{\rm tSZ}  E(z)^{-2/3} \left( \frac{D_{\rm A}(z)}{500\,{\rm Mpc}}\right)^2,
\label{eq:yintrinsic}
\end{equation}
$D_{\rm A}(z)$ being the angular distance to the QSO in Mpc.  Applying Eq.~(22) of~\cite{2010A&A...517A..92A}, we estimate the total mass of the halo as
%\begin{equation}
%\estMfive^{\alpha}= \left[ (1-b) \widehat{M}_{500,{\rm true}} \right ]^{\alpha} = \left [3 \times 10^{14} \left ( {h \over 0.7} \right ) ^{-1} \right ]^{\alpha}  {\widehat{A}_{\rm tSZ}  E(z)^{-2/3} \over A_{\rm x}(z)} M_{\odot}^{\alpha},
%\label{eq:mhalo}
%\end{equation}
\begin{equation}
\estMfive = (1-b) \widehat{M}_{500,{\rm true}} = 3 \times 10^{14} \left ( {h \over 0.7} \right ) ^{-1} {\left[{\widehat{\rm A}_{\rm tSZ}  E(z)^{-2/3} \over A_{\rm x}(z)}\right]}^{1/\alpha} M_{\odot} ,
\label{eq:mhalo}
\end{equation}
with $\alpha=1.78$ and $A_{\rm x}(z)=2.925 \times 10^{-5} \times 0.6145 \times \left ( {h \over 0.7} \right ) ^{-1} \times D_{\rm A}^{-2}(z) \times \left ( {\pi \over 180} {1 \over 60}  \right ) ^{-2} {\rm arcmin}^2$. The so-called mass bias parameter, $b$, accounts for any bias between the estimated mass and the true halo mass, $M_{500,{\rm true}}$, for example from violation of hydrostatic equilibrium and/or X-ray instrument calibration.  \Planck\ SZ cluster counts require high values of the mass bias, $(1-b)\sim 0.6-0.7$, for consistency with the base $\Lambda$CDM model favored by \Planck's measurements of the primary CMB anisotropies \citep[see][and references therein]{planck2013-XX, planck2015-XXIV}.\\
The $\estMfive - \widehat{Y}_{500}$ relation can be derived from the last two equations: 
\begin{equation}
\frac{\estMfive}{10^{13} \, M_{\odot}}=  \left( \frac{h}{0.7}\right)^{1/\alpha-1} \left[\frac{\widehat{Y}_{500} }{2.00 \times 10^{-6} \, {\rm arcmin}^2}\right]^{1/\alpha}.
\label{eq:ymhalo}
\end{equation}

\subsection{Statistics}
\subsubsection{Average values for the full QSO sample}
\label{sec:avg}
As described before, the individual QSO signals are expected to be faint, so we need to average them over the population.  Here, we detail the computation of this average.  We adopt the generic notation $\hat{w}_{\lambda}$ for $\widehat{A}_{\lambda}$, $\estMd$, $\estLs$, $\widehat{Y}_{500}$ or $ \widehat{F}_{i}$, arranging them into a vector $\hat{\vec{w}}$, and assume that they are Gaussian distributed. 

If these different components of $\hat{\vec{w}}$ are not correlated, we employ the usual inverse-variance weighted average:
\begin{equation}
\langle \hat{w}_{\rm \lambda} \rangle= \left[\sum_{p}{\frac{1}{\sigma^2[\left(\hat{w}_{\lambda}\right)_{p}]}}\right]^{-1} \left[\sum_{p}{\frac{\left(\hat{w}_{\lambda}\right)_{p}}{\sigma^2[\left(\hat{w}_{\lambda}\right)_{p}]}}\right],
\end{equation}
\begin{equation}
\sigma[\langle{\hat{w}_{\lambda}\rangle}]= \sqrt{\left[\sum_{p}{\frac{1}{\sigma^2[\left(\hat{w}_{\lambda}\right)_{p}]}}\right]^{-1}} ,
\end{equation}
with $p$ the index over the QSO catalogue.

In the correlated case, the average values are obtained according to
\begin{equation}
\label{eq:average_correlation}
\langle{\vec{\hat{w}} \rangle}= \left[\sum_{p}{\vec{C}^{-1}[\left(\hat{\vec{w}}\right)_{p}]}\right]^{-1} \cdot \left[\sum_{p}{\vec{C}^{-1}[\left(\hat{\vec{w}}\right)_{p}]  \cdot \left(\hat{\vec{w}}\right)_{p}} \right],
\end{equation}
\begin{equation}
\label{eq:sigma_average_correlation}
\vec{C}[\langle{\vec{\hat{w}} \rangle}]= \left[\sum_{p}{\vec{C}^{-1}[\left(\hat{\vec{w}}\right)_{p}]}\right]^{-1}.
\end{equation}

\subsubsection{$\chi^{2}$-test for the hypothesis}
\label{sec:chi2-test for the hypothesis}
If the assumptions on the nature and number of components are correct, the measured flux densities, $\widehat{F}_{i}$, and the reconstructed model flux densities, $\sum{\widehat{A}_{\lambda} \, \left({S}_{\lambda} \right)_{i} (z)}$, must be consistent within the uncertainties. We introduce the residual flux density for a given QSO as
\begin{equation}
R_{i}=\widehat{F}_{i} - \sum_{\lambda}{\widehat{A}_{\lambda}  \, \left(S_{\lambda} \right)_{i} (z)}.
\end{equation}
Correlation between the measured flux densities, $\widehat{\vec{F}}$, and the model amplitudes, $\vec{\widehat{A}}$, leads to the following expression for the variance of the residuals:
\begin{equation}
\begin{split}
C[\vec{R}]_{i j} &  \equiv  \langle{{R_{i}} {R_{j}} \rangle}  -\langle{{R_{i}} \rangle}\langle{{R_{j}} \rangle} \\
    & = C[\widehat{\vec{F}}]_{i j} - \sum_{\lambda, \mu}{ \left({S}_{\lambda}\right)_{i}(z) C[\vec{\widehat{A}}]_{\lambda, \mu} \left({S}_{\mu}\right)_{j}(z)},
    \end{split}
\end{equation}
where 
\begin{equation}
\begin{split}
C[\widehat{\vec{F}}]_{i j}  & \equiv  \langle{{\widehat{F}_{i}} {\widehat{F}_{j}} \rangle}- \langle{{\widehat{F}_{i}} \rangle}\langle{{\widehat{F}_{j}} \rangle} \\
    & =   N_{R_{500}} \left[\int d^2k\; 
     \frac{ \left({t_{\rm qso}^{*}}\right)_{i}(\vec{k}) \, \left({t_{\rm qso}}\right)_{j}(\vec{k}) \, {P_{i j}(\vec{k})} }{{P_{ii}(\vec{k})} \, {P_{jj}(\vec{k})}}\right]^{-1}. %\int d^2x \; \tau_{\rm qso}(\vec{x})
\end{split}
\end{equation}
with $N_{R_{500}} = \int_{r<R_{500}} dr \; 4\pi r^2 \tau_{\rm qso}(r)$. The residuals ${R}_{i}$ are expected to be Gaussian distributed with zero mean. We employ Eq.~(\ref{eq:average_correlation}) and~(\ref{eq:sigma_average_correlation}) with $\vec{R}$, instead of $\vec{\hat{w}}$, to compute the average value, and define the $\chi^{2}$ for the average residual as
\begin{equation}
\label{eq:chi2}
\chi^{2}= \langle{\vec{R} \rangle}^t \cdot \vec{C}^{-1}[\langle{\vec{R} \rangle}] \cdot \langle{\vec{R} \rangle}.
\end{equation}
The average residual vector $\langle{\vec{R} \rangle}$ has seven components, one per frequency. We expect the $\chi^{2}$ to follow Student's law with seven degrees of freedom, illustrated with simulations in~Sect.~\ref{sec:dof}.  The value of $\chi^{2}$ depends on the  number of components and their frequency dependance. In particular, the dust emission is parametrized by $T_{\rm d}$ and $\beta_{\rm d}$, so the $\chi^{2}$ also depends on their values. If we leave these two parameters free, the $\chi^{2}$ would be expected to follow Student's law with 7-2=5 degrees-of-freedom.

We study the average residual vector, $\langle{\vec{R} \rangle}$, because it is the most useful quantity for identifying deviations between observed signals and model predictions (dust, dust+tSZ, dust+synchrotron or dust+tSZ+synchrotron).  Individual $\chi_i^2$ obtained from the residuals ${R}_{i}$ can also be computed, but given the large measurement uncertainties, study of the individual $\chi_i^2$ distributions do not efficiently discriminate models.

\subsection{Marginalization over dust properties}
\label{sec:Marginalization on the (T,beta) plane} 
The previous section considered computation of $\langle{{\hat{w}_{\lambda}} \rangle}$ and $\chi^{2}$ for fixed dust properties, described by the parameters $T_{\rm d}$ and $\beta_{\rm d}$. Although $\widehat{A}_{\rm dust}$ is only weakly dependent on these parameters, the derived dust mass, $\estMd$ (Eq.~\ref{eq:mdust}), is more sensitive to them.  We use Powell's algorithm to minimize the $\chi^2$ relative to $T_{\rm d}$, $\beta_{\rm d}$ to find their best-fit values.  We also marginalize over $T_{\rm d}$, $\beta_{\rm d}$ when determining the physical properties of the QSO environment, $\estMd$, $\estLs$ and $\widehat{Y}_{500} $.

By ``data'' in the following, we will mean the combination of the \Planck\ maps and the BOSS QSO positions. The $\langle{{\hat{w}_{\lambda}} \rangle}$ being Gaussian distributed (Sec.~\ref{sec:avg}), we write
\begin{equation}
P({\hat{w}}_{\lambda} | T_{d}, \beta_{d}, {\rm data}) \propto \exp\left(-\frac{1}{2} \left(\frac{{\hat{w}}_{\lambda}-\langle{{\hat{w}_{\lambda}} \rangle}}{\sigma[\langle{\hat{{w}}_{\lambda}\rangle}]}
\right)^2\right).
\end{equation}
With $P({\rm data} | T_{\rm d}, \beta_{\rm d})  \propto  \exp\left(-\frac{\chi^2}{2}\right)$, Bayes' theorem gives 
\begin{equation}
P(T_{\rm d}, \beta_{\rm d} | {\rm data})  \propto  P(T_{\rm d}, \beta_{\rm d}) \times \exp\left(-\frac{\chi^2}{2}\right).
\end{equation}
Assuming a flat prior on $P(T_{\rm d}, \beta_{\rm d}) $, we compute the probability of ${\hat{w}}_{\lambda}$, $T_{\rm d}$, and $\beta_{\rm d}$ given the data as,
\begin{equation}
P({\hat{w}}_{\lambda}, T_{\rm d}, \beta_{\rm d} | {\rm data}) \propto P({\hat{w}}_{\lambda} | T_{\rm d}, \beta_{\rm d}, {\rm data}) \times P(T_{\rm d}, \beta_{\rm d} | {\rm data}).
\end{equation}

Averages are now calculated by marginalizing over $T_{\rm d}$ and $\beta_{\rm d}$ and via an integration over ${\hat{w}}_{\lambda}$:
\begin{equation}
\overline{{w}}_{\lambda}=  \iiint {\hat{w}}_{\lambda} P({\hat{w}}_{\lambda}, T_{\rm d}, \beta_{\rm d} | {\rm data}) dT_{\rm d} d\beta_{\rm d} {\hat{w}}_{\lambda},
\end{equation}
with variance 
\begin{equation}
\sigma^2[{\overline{w}}_{\lambda}]=  \iiint ({\hat{w}}_{\lambda}-\overline{{w}}_{\lambda})^2 P({\hat{w}}_{\lambda}, T_{\rm d}, \beta_{\rm d} | {\rm data})  dT_{\rm d} d\beta_{\rm d} d{\hat{w}}_{\lambda}.
\end{equation}

%We note the quantity marginalized over $\rm T_{\rm d}$ and $\rm \beta_{\rm d}$ with an over-line. Since the single frequency flux does not depend on the choice of $\rm T_{\rm d}$ and $\rm \beta_{\rm d}$, the notations $\langle\widehat{\vec{F}}\rangle$ (average over the QSO population) and $\overline{\vec{F}}$ are equivalent.

\section{Simulations}
\label{sec:simu}
We validate our methods by injecting simulated QSOs into the \Planck\ maps and re-extracting their properties.  The simulations are described in Sect.~\ref{sec:mocks}.  Section~\ref{sec:dof} shows that our $\chi^{2}$ statistic carries the expected number of degrees-of-freedom.  We demonstrate that the MMFs properly recover the properties of the mock catalogs in Sect.~\ref{sec:simures}, and in Sect.~\ref{sec:syst} we examine the sensitivity our results to the dust model.

\subsection{Mock catalogs and injected maps}
\label{sec:mocks}
We build mock catalogs directly from the original BOSS sample, keeping the same QSO redshift distribution (Fig.~\ref{fig:nz_qso}) but drawing the sky positions at random outside the \Planck\ point-source mask.  We fix the QSO host halo mass $M_{500}=2.7 \times 10^{13} M_\odot$, close to the value found from the data in Sect.~\ref{sec:hotgas}.  We then compute the expected tSZ flux for each host from $(z,M_{500})$ using Eq.~(\ref{eq:mhalo}). 

We assign to each host halo a constant dust mass $(M_{\rm dust})_{\rm input}=2.5 \times 10^8 M_\odot$ emitting with a modified blackbody spectrum at $(T_{\rm d})_{\rm input}=25 \, {\rm K}$, $(\beta_{\rm d})_{\rm input}=2.5$.  When specified, we also implement variations in the dust temperature $(T_{\rm d})_{\rm input}$ using a Gaussian distribution with mean $(T_{\rm d})_{\rm input}=25 \, {\rm K}$ and standard deviation $(\sigma_{\rm T})_{\rm input} =5 {\rm K}$.

With this model we compute the signal of each QSO in the \Planck\ bands (from tSZ and dust) and we inject them directly into the maps using our source profile, $\tau_{ \rm qso}$, convolved with the individual channel beams.  We artificially fix $\theta_{\rm s}$ to 0.27 arcmin for the injected profile to avoid possible biases due to the mismatch between the extraction profile and the halo profile at low redshift. The sensitivity of our result to this assumption is studied in Sect.~\ref{sec:syssize}.

\subsection{Degrees of freedom}
\label{sec:dof}
\begin{figure}
\centering
\includegraphics[width=.5\textwidth]{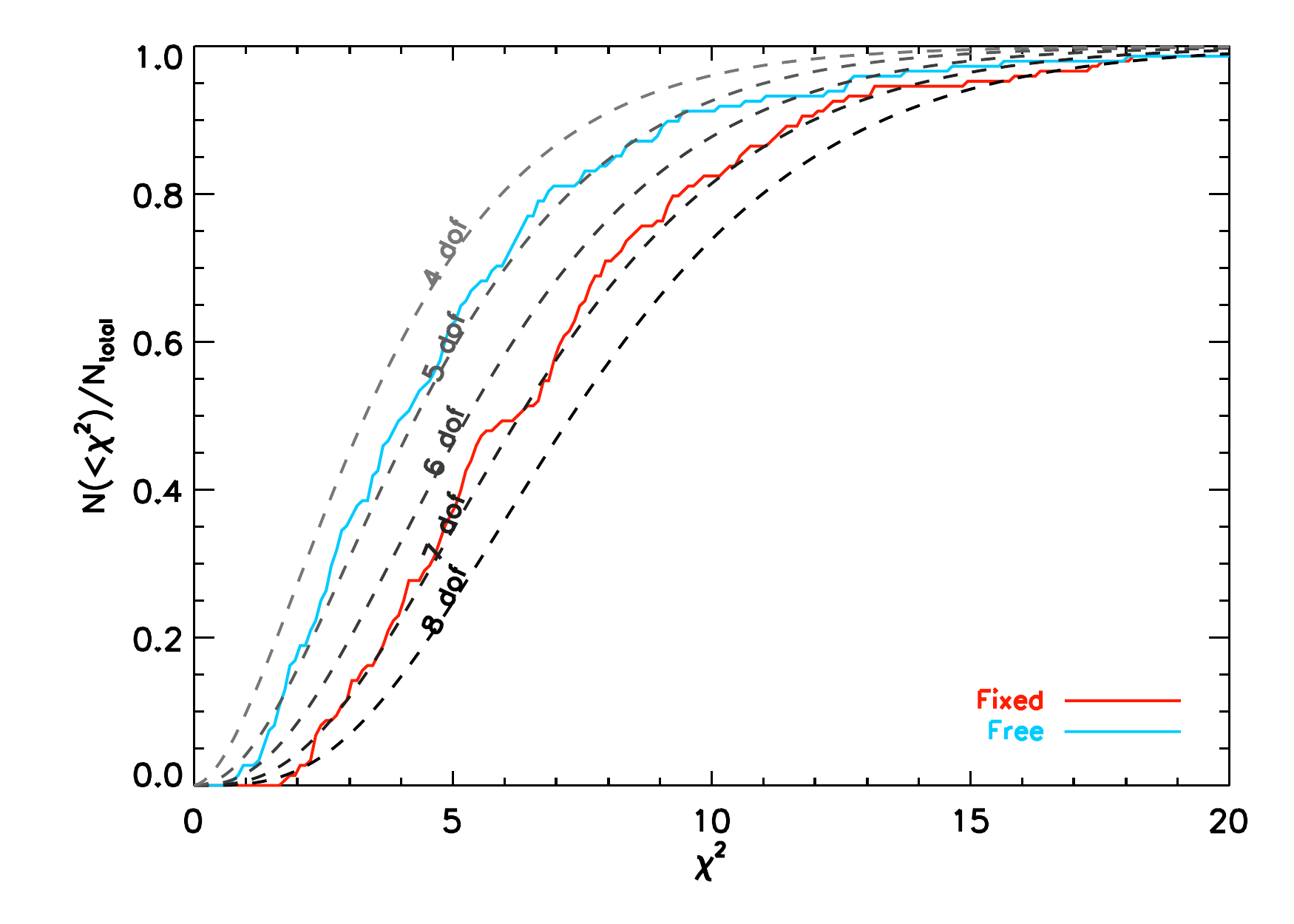}
\includegraphics[width=.5\textwidth]{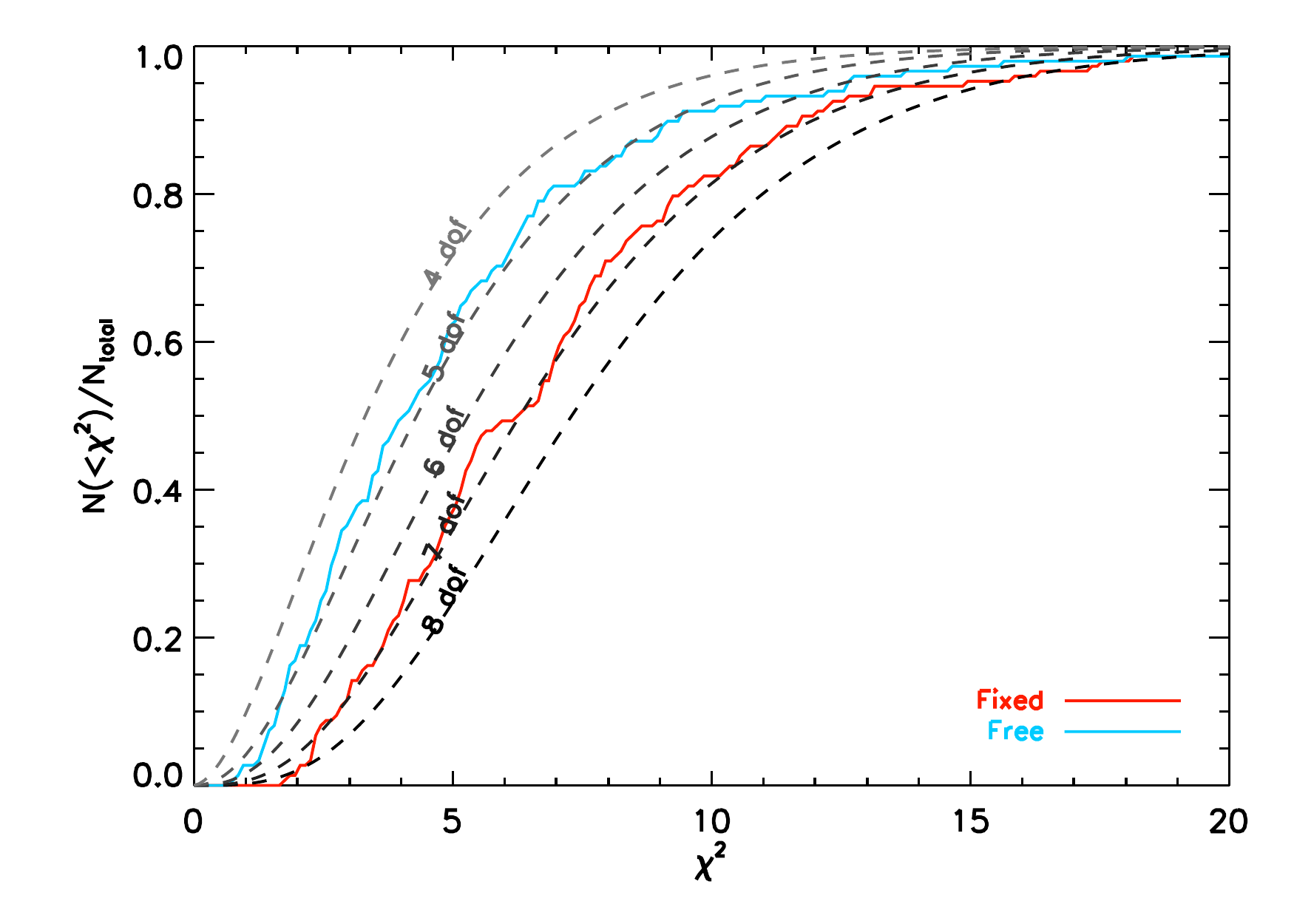}
\caption{Cumulative $\chi^2$  distributions when $T_{\rm d}$ and $\beta_{\rm d}$ are fixed in the filtering (solid red curve) or when they are left free (solid blue curve). \textit{Top panel:}  Distribution for a dust-only filtering on a dust-only mock catalogue. \textit{Bottom panel:}  Distribution for a dust+tSZ filtering on a dust+tSZ mock catalogue. In both cases (dust-only or dust+tSZ), fixing $T_{\rm d}$ and $\beta_{\rm d}$ leads to a $\chi^2$ distribution with seven degrees-of-freedom (corresponding to our seven maps), while leaving them free reduces the number of degrees-of-freedom to five.}
\label{fig:distrib_chi2}
\end{figure}

We test the number of degrees-of-freedom for the $\chi^2$ statistic defined in Sect.~\ref{sec:chi2-test for the hypothesis} by injecting into the \Planck\ maps a mock catalogue with only dust emission from the QSOs and then extracting the signal with the dust-only MMF at fixed $T_{\rm d}=(T_{\rm d})_{\rm input}$ and $\beta_{\rm d}=(\beta_{\rm d})_{\rm input}$.  We bin our catalogue into 148 sub-catalogs of 2000 QSOs each to compute 148 independent $\chi^2$ values.  The cumulative distribution of these values is given in the top panel of Fig.~\ref{fig:distrib_chi2} as the solid red line.  Student's cumulative distribution with varying degrees-of-freedom (dof) are shown as the dashed lines. The red line follows Student's law with seven degrees-of-freedom, corresponding to the seven \Planck\ maps.  When leaving $T_{\rm d}$ and $\beta_{\rm d}$ free and choosing their best-fit values as the point  minimizing the $\chi^2$, we obtain the solid blue line with 7-2=5 degrees-of-freedom, as expected.

We next inject a mock catalogue with both dust and tSZ emission and re-extract the QSO signals using a dust+tSZ filter to compute the $\chi^2$. The results are shown in the bottom panel of Fig.~\ref{fig:distrib_chi2}.  Fixing $T_{\rm d}$ and $\beta_{\rm d}$ to the input values also leads to a $\chi^2$ with seven dof. Leaving the two dust parameters free lowers the $\chi^2$ to five dof, as for the dust-only case.
Although we are searching for an additional component in the data between the first and the second tests, the $\chi^2$ conserves 7/5 dof for ($T_{\rm d}$, $\beta_{\rm d}$) fixed/free respectively.  This is because the correlations between our residuals are taken into account in the covariance matrix $\vec{C}[\langle{\vec{R} \rangle}]$ in~Eq.~(\ref{eq:chi2}). In other words, $\vec{C}[\langle{\vec{R} \rangle}]$ changes when considering the dust-only MMF or the dust+tSZ MMF, so the $\chi^2$ statistic defined in~Eq.~(\ref{eq:chi2}) does not depend on the number of components assumed for the MMF.

\subsection{Extraction of simulated QSO}
\label{sec:simures}
We first consider null tests where the signals are extracted at random positions in the \Planck\ maps. Table~\ref{tab:nulltest} shows the output for three filters (dust, dust+tSZ, dust+synch) assuming fixed values of $T_{\rm d}=25$\,K and $\beta_{\rm d}=2.5$ in the extraction. The output dust masses, SZ masses and synchrotron luminosities are compatible with zero, as expected, and the $\chi^{2}$ do not show any significant deviation from Student's law with the expected seven dof.

Results of the extraction on mock catalogs injected into the \Planck\ maps are summarized in Table~\ref{tab:simulation}. The recovered dust masses, SZ masses, and synchrotron luminosities are marginalized over $T_{\rm d}$ and $\beta_{\rm d}$ as described in Sect.~\ref{sec:Marginalization on the (T,beta) plane}. 

For the injected dust+tSZ catalogue, the dust+tSZ filter recovers a unbiased estimate of the input dust temperature and spectral index. The input dust mass and the halo mass from the tSZ are also recovered at $S/N \sim 8$ and $S/N \sim 14$.  The dust mass recovered using the dust-only filter is biased by $+3.4 \sigma$ due to contamination of the dust emission by the tSZ. The recovered dust spectral index is also biased by $-3.7 \sigma$.  Adopting the dust+synch filter leads to a significant detection of negative synchrotron luminosity at $+4.6 \sigma$, which mimics the tSZ effect at low frequencies.  

\begin{figure}
\centering
\includegraphics[width=.5\textwidth]{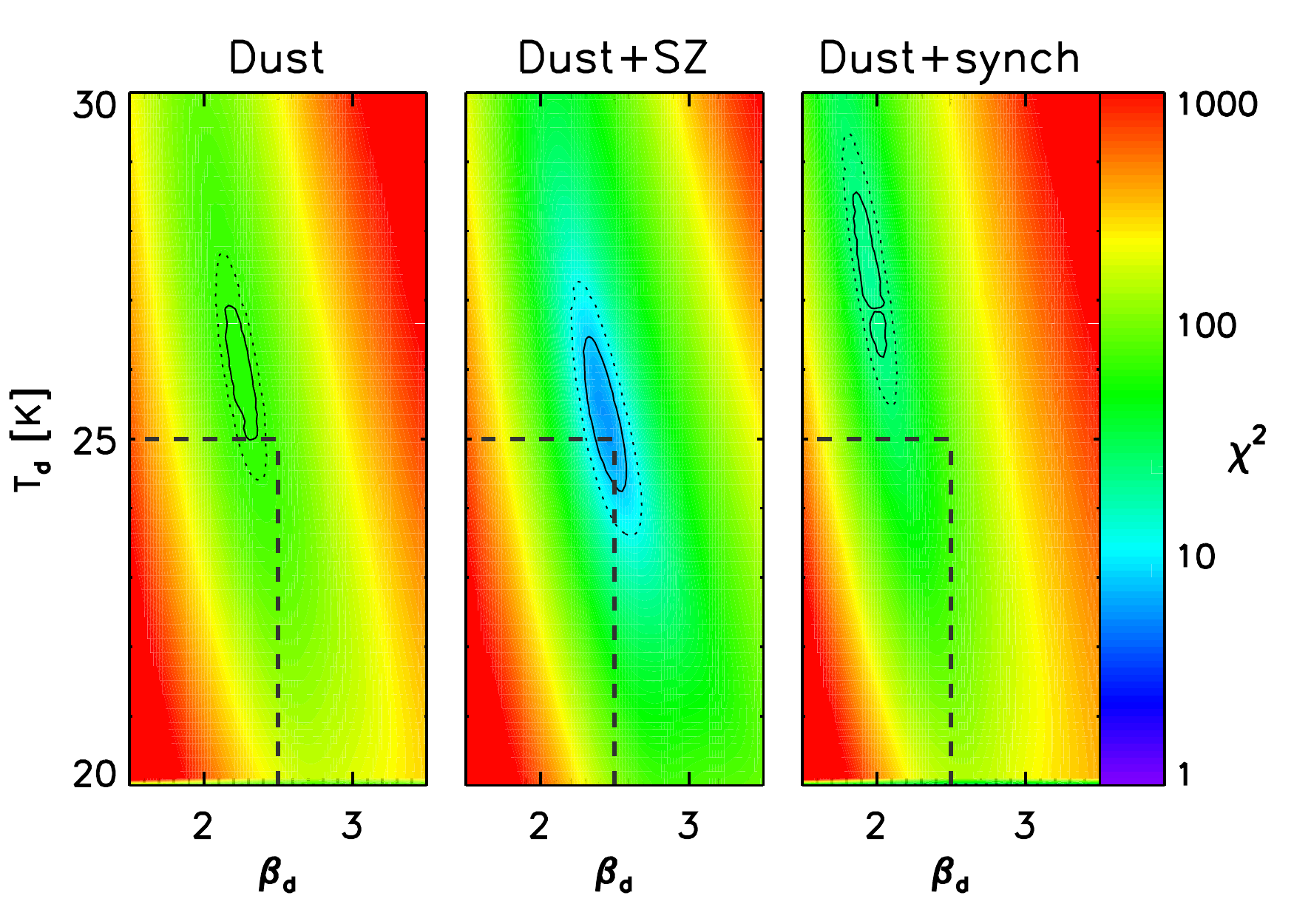}
\caption{Contours of $\chi^{2}$ for the mock QSO dust+tSZ catalogue using the dust, the dust+tSZ, and the dust+synch filters. The solid black line indicates the 1$\sigma$ deviation from the minimal $\chi^2$ and the dotted black line the 2$\sigma$ deviation. The horizontal and vertical dashed lines mark the position of the input $\rm T_{\rm d}$ and $\beta_{\rm d}$.}
\label{fig:maps_chi2_beta_Tdust}
\end{figure}

The dust-only extraction provides a reduced $\chi^2/5$ value of 12.1, larger than the dust+tSZ value of 1.1, indicating that the dust+tSZ model must be preferred over the dust-only model. The dust+synch model provides a better $\chi^2/5$ than the dust-only case (4.0), but the significantly negative value for the synchrotron luminosity discards the model. The ($T_{\rm d}$,$\beta_{\rm d}$) contours for the three filters are displayed in Fig.~\ref{fig:maps_chi2_beta_Tdust}. The dust+tSZ filter shows contours enclosing the input values while the two other filters are biased.

We also implemented the triple dust+tSZ+synch filter and applied it to the dust+tSZ simulation. The triple filter transforms part of the tSZ emission into a negative (i.e., unphysical) synchrotron emission, as shown in Table~\ref{tab:simulation}. It recovers biased $T_{\rm d}$, $\beta_{\rm d}$, and $M_{\rm dust}$. The reduced $\chi^2/5$ increases with respect to the dust+tSZ filter, showing that the dust+tSZ fit must be preferred over the dust+tSZ+synch fit. We show in Sect.~\ref{sec:results} that the triple filter exhibits the same behavior on the data. We will thus present our main results using the dust, the dust+tSZ, and the dust+synch filters, depending on the value of the reduced $\chi^2/5$.

%\begin{table}
\begin{table*} %referee
		%\hspace{-0.5 cm}
		\centering
		\caption{Null test performed at fixed ($T_{\rm d}$=25 K, $\beta_{\rm d}$=2.5). The filtering is done at random positions in the \Planck\ maps.}
		\label{tab:nulltest}
	%\begin{tabular}{@{\hspace{-0.25 cm}}lcccccc}
	\begin{tabular}{lcccccccc}
		\hline
		\hline
		Filter & $\chi^{2}$/7 &$T_{\rm d}$  & $\beta_{\rm d}$  &  $\langle{\widehat{M}_{\rm dust}\rangle} $ & $\langle{\widehat{Y}_{\rm 500}\rangle} $& $\langle{\widehat{L}_{\rm synch}\rangle}$   \\
		  & & (K)  &  & ($10^{8} \, M_\odot$) & $(10^{-6} \, arcmin^2)$ &  ($10^{-3} \, L_\odot {\rm Hz^{-1}}$) \\
		\hline
		\multicolumn{7}{l}{Mock catalogue : \textit{Null test}} \\
		{\it input} & - & - & - &  0 & 0 & 0\\
		\hline
		dust & 0.7 & 25 & 2.5 &$ 0.003  \pm  0.012$ & - & - \\
		dust+tSZ & 1.2 & 25 & 2.5 &$0.003 \pm  0.012$ & $0.77 \pm 0.78$  & -  \\
		dust+synch & 3.5 & 25 & 2.5  & $0.003 \pm  0.012$ &  - &  $-0.05 \pm 0.13$   \\
		\hline
	\end{tabular}
%\end{table}
\end{table*} %referee

\begin{table*}
		%\hspace{-0.5 cm}
		\centering
		\caption{Extracted QSO properties for the mock catalogs, averaged over ($T_{\rm d}$, $\beta_{\rm d}$) as described in Sec.~\ref{sec:Marginalization on the (T,beta) plane}. The first mock catalogue (upper part of the Table) features both the tSZ and the dust emission. The second catalogue (bottom part) also includes variation in the dust temperature $T_{d}$. For each case the $\it input$ line gives the injected values. The other lines shows the values extracted with the various filters. For each simulation, the filter assuming the same components as the input is highlighted in gray.}
		\label{tab:simulation}
	%\begin{tabular}{@{\hspace{-0.25 cm}}lcccccc}
	\begin{tabular}{lcccccccc}
		\hline
		\hline
		Filter & $\chi^{2}$/5 &$T_{\rm d}$  & $\beta_{\rm d}$  &  $\overline{M_{\rm dust}} $ & $\overline{Y_{500}} $& $\overline{M_{500}}$& $\overline{L_{\rm synch}}$   \\
		  & & (K)  &  & ($10^{8} \, M_\odot$) & $(10^{-6} \, arcmin^2)$ &  ($10^{13} \, M_\odot$) & ($10^{-3} \, L_\odot {\rm Hz^{-1}}$) \\
		\hline
		\multicolumn{7}{l}{Mock catalogue : \textit{dust+tSZ}} \\
		{\it input} & - & 25 & 2.5  & 2.50 & 12.02 & 2.74 & 0\\
		\hline
		dust & 12.1 & $26.1 \pm 0.6$ & $2.24 \pm 0.07$  &   $3.54 \pm  0.31$ &  - &- & - \\
\rowcolor{lightgray}	dust+tSZ & 1.1 & $25.4 \pm 0.8$ & $2.43 \pm 0.1$ & $2.71 \pm  0.34$ &  $13.03\pm 1.63$& $2.86 \pm 0.20$  &  -  \\
		dust+synch & 4.0 & $27.4 \pm 0.8$ & $1.95 \pm 0.08$ &   $5.18 \pm  0.42$ & - &- & $-0.65 \pm 0.14$  \\
		dust+tSZ+synch & 2.9 & $27.4 \pm 1.0$ & $2.05 \pm 0.13$ &   $4.29 \pm  0.65$ & $8.68\pm 1.90$ & $2.27 \pm 0.30$& $-0.37 \pm 0.17$  \\
		\hline
		\multicolumn{7}{l}{Mock catalogue : \textit{dust+tSZ with $T_{\rm d}$ Gaussian distributed}} \\
		{\it input} & - & $25 \pm 5$ & 2.5 & 2.50 & 12.02 & 2.74 & 0\\
		\hline
		dust & 12.7 & $30.9 \pm 1.2$ & $2.06 \pm 0.08$  &   $3.27 \pm  0.22$ & - &- & -  \\
		dust+tSZ & 1.1 & $29.8 \pm 0.8$ & $2.27 \pm 0.08$ &  $2.46 \pm  0.27$ & $13.03 \pm 1.70$  & $2.86 \pm 0.21$ & - \\
		dust+synch & 3.7 & $33.6 \pm 0.4$ & $1.75 \pm 0.05$ &  $4.69 \pm  0.29$ & - & - & $-0.69 \pm 0.13$  \\
		dust($\sigma_{T}$=2 K)+tSZ & 1.1 & $29.3 \pm 0.8$ & $2.28 \pm 0.08$ &  $2.47  \pm  0.27$ & $12.81 \pm 1.68$  & $2.83 \pm 0.21$ & - \\
\rowcolor{lightgray}		dust($\sigma_{T}$=5 K)+tSZ & 1.1 & $26.4 \pm 1.2$ & $2.37 \pm 0.10$  &  $2.67 \pm  0.29$ & $12.97 \pm 1.68$  & $2.85 \pm 0.21$ & - \\
		dust($\sigma_{T}$=10 K)+tSZ  & 1.2 & $7.5 \pm 1.6$& $2.95 \pm 0.08$  &  $5.43 \pm  0.40$ & $11.45 \pm 1.31$  & $2.66 \pm 0.17$ & - \\
		\hline
	\end{tabular}
\end{table*}

\begin{figure}
\centering
\includegraphics[width=.5\textwidth]{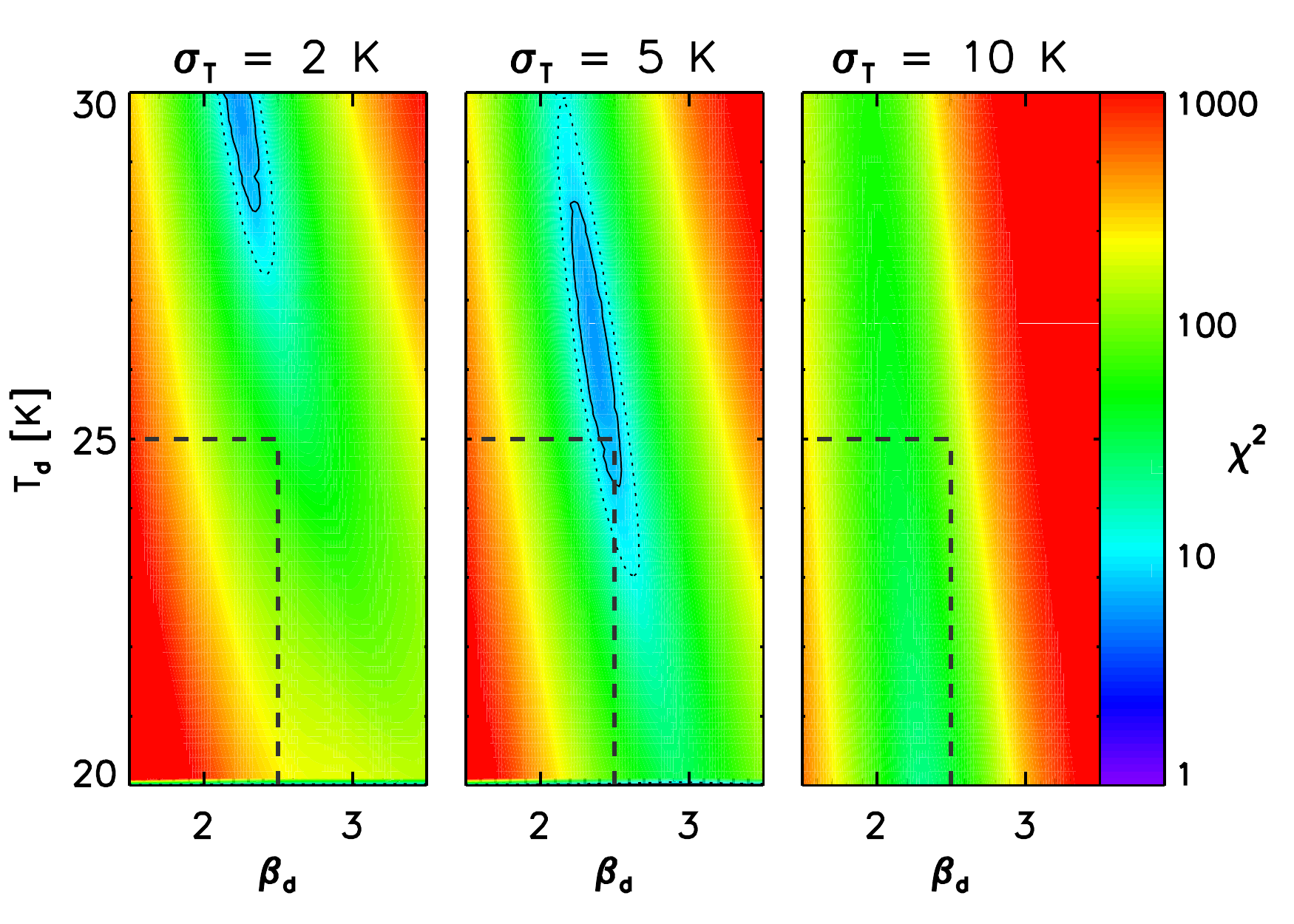}
\caption{Contours of $\chi^{2}$ for the mock QSO dust ($\sigma_{\rm T}$=5 K)+tSZ catalogue, obtained using the dust($\sigma_{\rm T}$ =2\,K)+tSZ, the dust ($\sigma_{\rm T}$ =5\,K)+tSZ, and the dust ($\sigma_{\rm T}$ =10\,K)+tSZ filters. The solid black line indicates the 1$\sigma$ deviation from the minimal $\chi^2$, and the dotted black line the 2$\sigma$ deviation. The horizontal and vertical dashed lines mark the position of the input $\rm T_{\rm d}$ and $\beta_{\rm d}$.}
\label{fig:maps_chi2_beta_Tdustgauss}
\end{figure}
 
\subsection{Possible systematics due to the dust characteristics}
\label{sec:syst}
We test the sensitivity of the extraction method to improper modeling of the dust by injecting mock QSOs with a dust temperature following a Gaussian distribution, described in Sect.~\ref{sec:mocks}, and then recovering the parameters with the different MMF. Results are shown in the bottom part of Table~\ref{tab:simulation}. The dust ($\sigma_{\rm T}=x$K)+tSZ filters are a dust+tSZ filter for which a Gaussian scatter of $x$ Kelvin in the dust temperature is added. As expected, the dust ($\sigma_{\rm T}$ =5\,K)+tSZ filter provides an unbiased estimate of the parameters. The derived dust and SZ masses are not very sensitive to the dispersion in the dust temperature: the dust+tSZ filter returns a satisfactory estimate of the two quantities. However, for this latter filter the recovered dust parameters $T_{\rm d}$ and $\beta_{\rm d}$ are significantly biased. This sensitivity is illustrated in Fig.~\ref{fig:maps_chi2_beta_Tdustgauss} displaying the ($T_{\rm d}$,$\beta_{\rm d}$) contours for the three last filters of Table~\ref{tab:simulation}.

\section{Results}
\label{sec:results}
We apply the method described in Sect.~\ref{sec:analysis} to the real \Planck\ data.  In Sect.~\ref{sec: Halo flux at Planck frequencies}, we show the average total flux of the QSO sample across the \Planck\ channels. We then estimate the contribution of the various components in Sect.~\ref{sec:dust} (dust), \ref{sec:synch} (synchrotron), and \ref{sec:hotgas} (hot gas).

\subsection{The QSO signal at \Planck\ frequencies}
\label{sec: Halo flux at Planck frequencies}
\begin{figure}
\centering
\includegraphics[width=.5\textwidth]{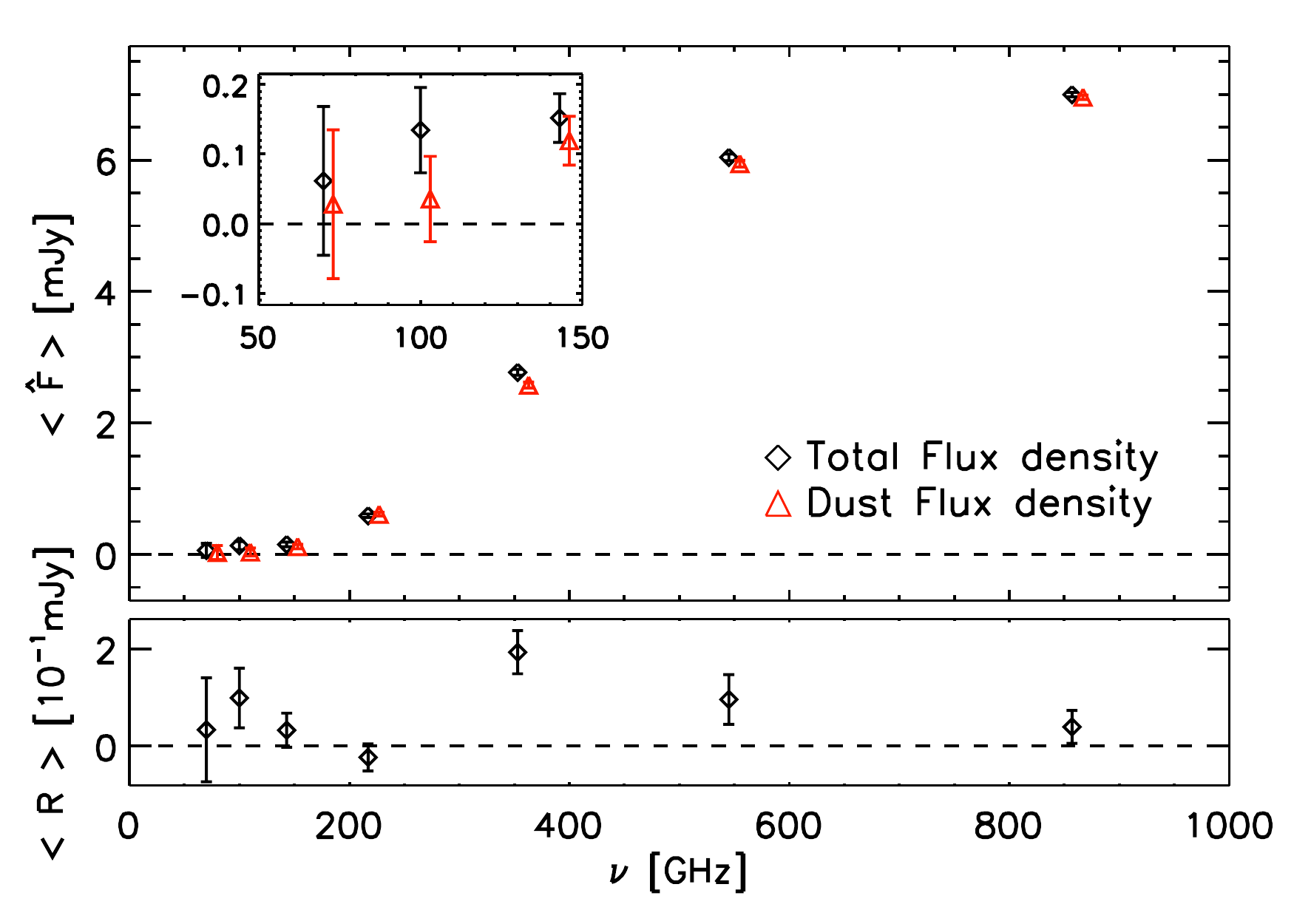}
\caption{Average flux density for the QSO population across \Planck\ channels (black diamonds), and dust signal estimated with the dust-only filter (red triangles) for $T_{\rm d}$ and $\beta_{\rm d}$ at the minimum $\chi^{2}$. The bottom panel shows the average residuals.}
\label{fig:flux_dust_all}
\end{figure}

We first estimate the mean flux density, $\langle\widehat{\vec{F}}\rangle$, of the 291 165 QSOs selected in Sect.~\ref{BOSS quasars} by averaging the individual flux densities in each band as extracted with a single frequency matched filter centered on the QSO positions.  The average total flux is shown as the black diamonds in Fig.~\ref{fig:flux_dust_all} (main panel and inset).  It is {\em strictly positive} in all \Planck\ bands above 100\,GHz, and positive but compatible with zero at 70\,GHz.  The signal continuously increases from the lowest to the highest frequencies: emission in the direction of the QSO is dominated by dust. Since none of the frequencies below 217\,GHz presents a negative flux, there is no obvious indication of strong tSZ emission.

Using the dust-only MMF with $T_{\rm d}$ and $\beta_{\rm d}$ adjusted to minimize the $\chi^2$ defined in Eq.~(\ref{eq:chi2}) ($T_{\rm d}=18.7 \, {\rm K}$, $\beta_{\rm d}=2.79$), we compute the mean dust flux density $\widehat{A}_{\rm dust}$ and plot the resulting mean dust spectrum as the red triangles in Fig.~\ref{fig:flux_dust_all}.  Dust accounts for essentially all the observed signal.  The residual (difference between the black diamonds and the red triangles) is shown in the bottom panel of the figure, and is consistent with zero except at 353\,GHz. This deviation explains why the reduced $\chi^2/5$ highlighted in the first gray line of Table~\ref{tab:DR12_data} is not good ($\chi^2/5$= 5.2). 

%We now use the dust-only MMF and leave $T_d$ and $\beta_d$ free. We keep $T_d$ and $\beta_d$ which minimizes the $\chi^2$ ($T_d=18.7 \, {\rm K}$, $\beta_d=2.79$) and compute the corresponding dust flux with the MMF. It is shown as red triangles in Fig.~\ref{fig:flux_dust_all}. The dust flux is a close match to the total flux. The residual (difference between the black diamonds and the red triangles) is shown in the bottom panel of the figure. It is consistent with zero, except at 353~GHz. This deviation explains why the reduced $\chi^2/5$ highlighted in the first gray line of Table~\ref{tab:DR12_data} is not good ($\chi^2/5$= 5.2). 

By marginalizing over $T_{\rm d}$ and $\beta_{\rm d}$, we obtain an average dust mass, $\overline{M_{\rm dust}}=(0.84 \pm 0.07)  \times 10^{8} M_\odot$. This value is $\sim 25$ times smaller than the estimated dust mass in optically-selected clusters, around $2 \times 10^{9} M_\odot$ \citep{2014A&A...571A..66G}. It is also five times smaller than the lowest estimated QSO dust mass of the \cite{2006ApJ...642..694B} sample. We note, however, that our average mass is dominated by low redshift objects ($z \sim 0.5$), as shown in Fig.~\ref{fig:dust_mass_all}, while the QSO sample studied in~\cite{2006ApJ...642..694B} resides at higher redshift. 

The recovered dust mass also depends strongly on the assumed value for $T_{\rm d}$ and $\beta_{\rm d}$. If we fix $\beta_{\rm d}=1.6$, as in \cite{2006ApJ...642..694B}, we find $\overline{M_{\rm dust}}=(1.70 \pm 0.08)  \times 10^{8} M_\odot$, in better agreement with the above-cited study, but the $\chi^2/5$ increases to 16.3, as shown in Table~\ref{tab:DR12_data}. This result demonstrates that the bulk of the BOSS QSO population contains significantly less dust than QSOs examined in these previous studies.

For the dust temperature, we find a marginalized $T_{\rm d} =  19.1 \pm 0.8$\,K, close to the dust temperature of normal galaxies~\citep[e.g.,][]{clemens2013}.  This is significantly lower than the temperature found by~\cite{2006ApJ...642..694B} and~\cite{2012ApJ...753...33D}, ranging between 33 and 55\,K, and 18.1 and 79.7\,K respectively. We note that \cite{2012ApJ...753...33D} use the two component model from \cite{2003MNRAS.338..733B} with $\beta_{\rm d}=2$, and \cite{2006ApJ...642..694B} fix $\beta_{\rm d}=1.6$ for the majority of their QSOs. We find $\beta_{\rm d} =  2.71 \pm 0.13$,  significantly higher than these values. When fixing $\beta_{\rm d}=1.6$ in our analysis, our result moves along the degeneracy line in the ($\rm T_{\rm d}$,$\rm \beta_{\rm d}$)-plane to reach $T_{\rm d}=28.2\pm 0.5$\,K, in better agreement with the values published in the previously mentioned analyses.  Restricting this test to high redshift QSOs leaves our conclusions unchanged (line 6 of Table~\ref{tab:DR12_data}).

\subsection{Redshift dependence of the dust properties}
\label{sec:dust}
In order to adapt the MMF to a possible variation of the QSO dust properties across redshift, we divide our QSO sample into nine regular bins of size $\Delta z=0.5$, shown as the vertical dashed lines in Fig.~\ref{fig:nz_qso}.  We include in the last bin the QSOs with $4<z<5$ because there are not enough statistics to fill a tenth bin.

\begin{figure}[!h]
\centering
\includegraphics[width=.5\textwidth]{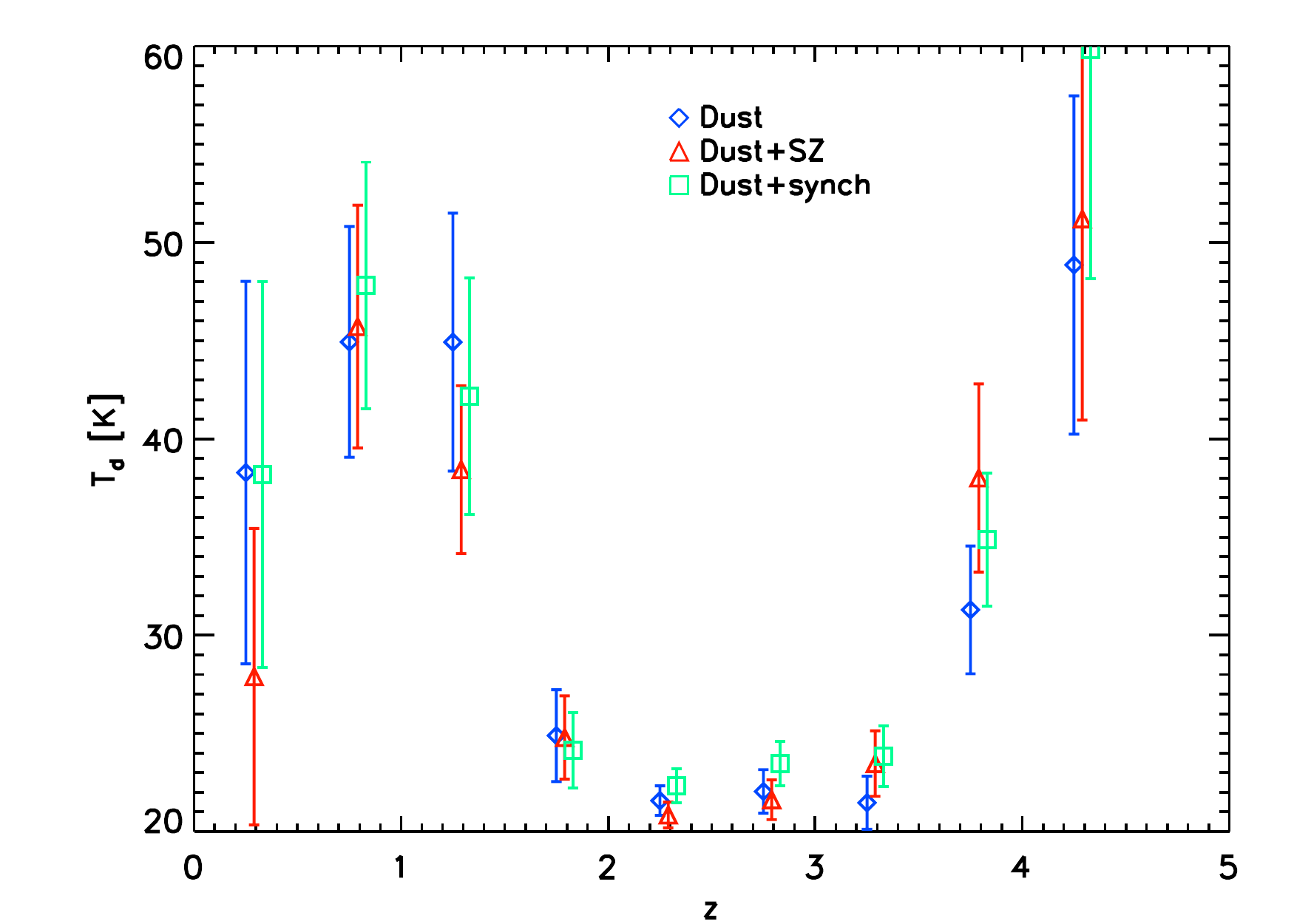}
\includegraphics[width=.5\textwidth]{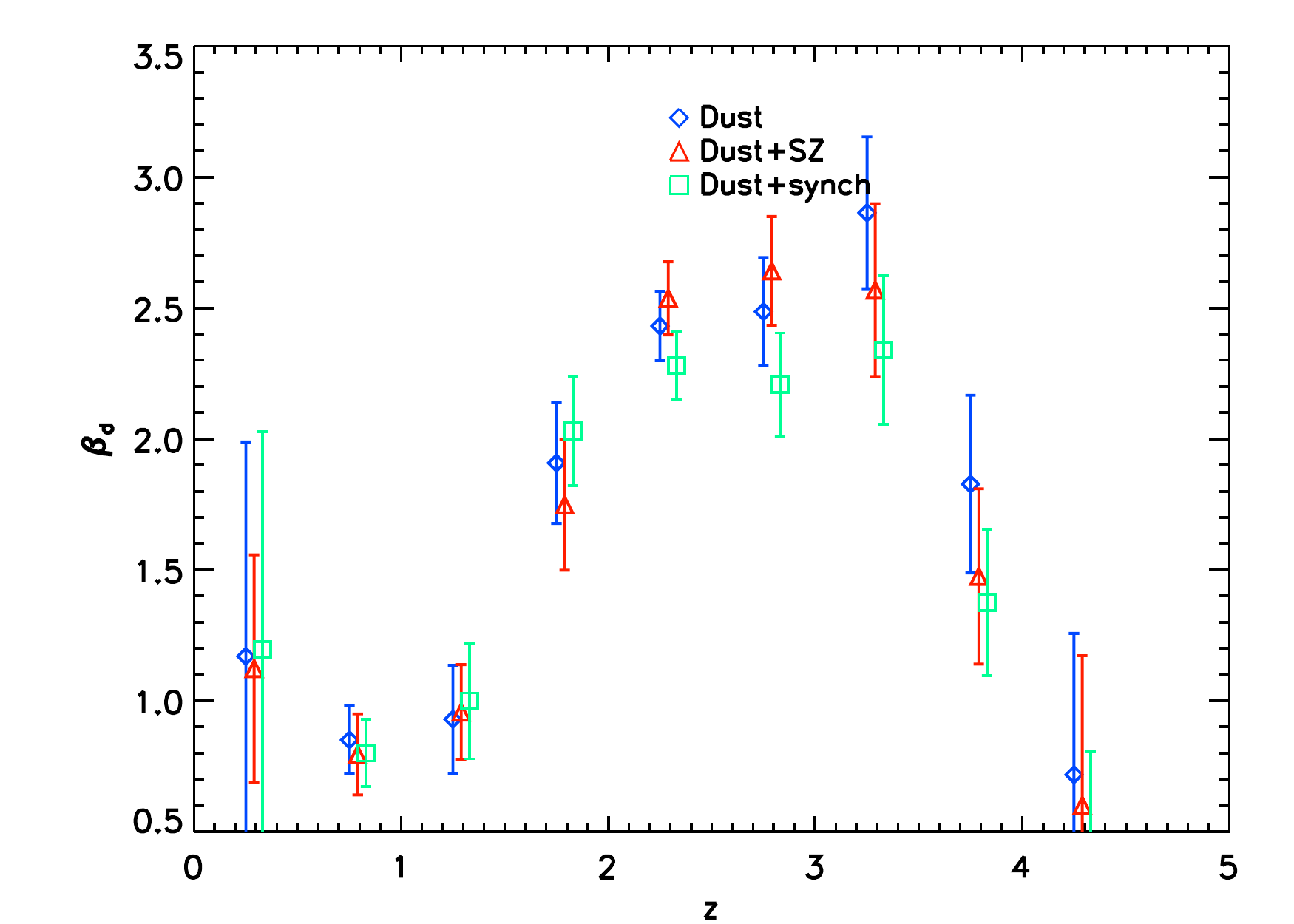}
\includegraphics[width=.5\textwidth]{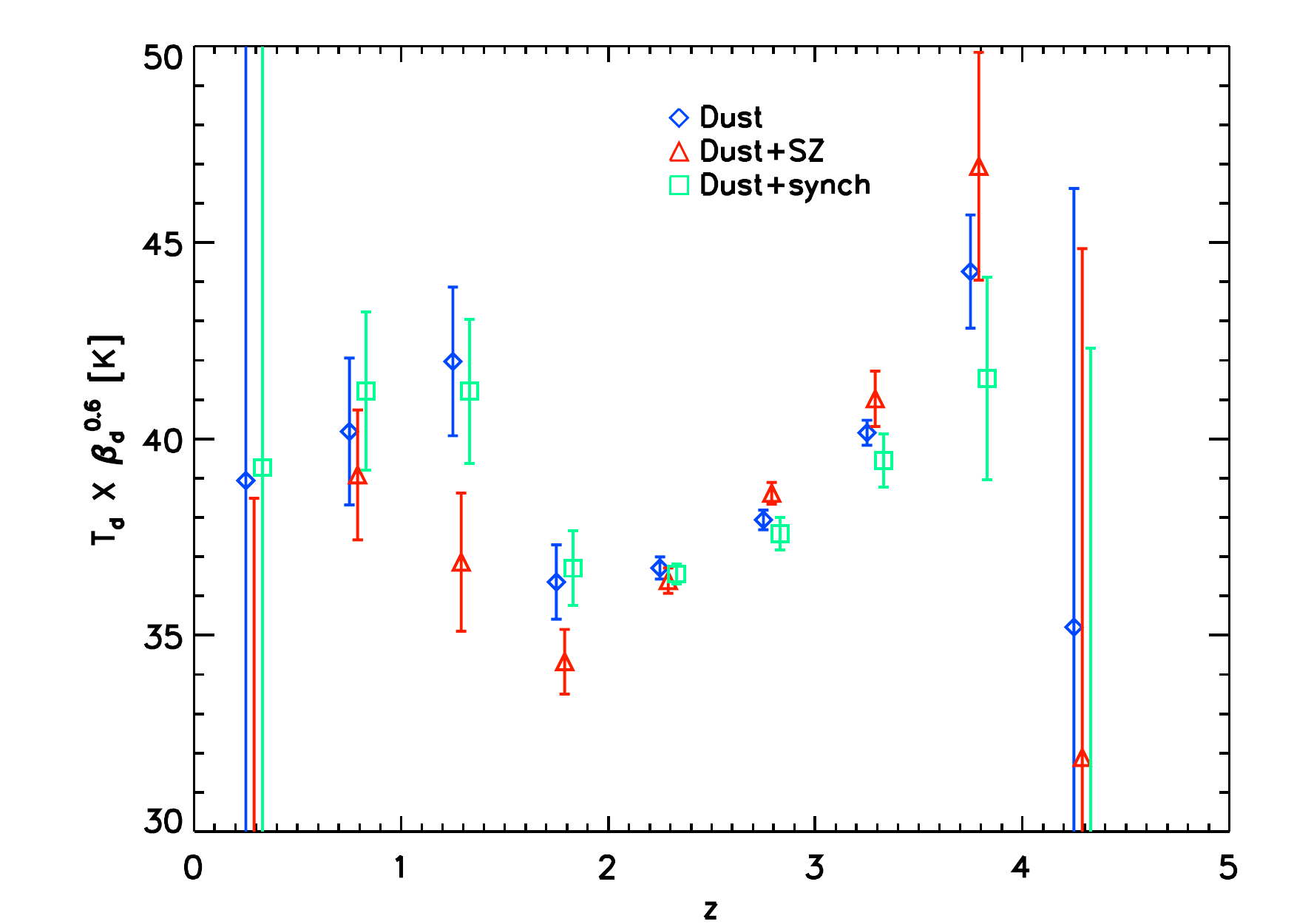}
%\caption{Average dust $\rm T_{\rm d} \times \rm \beta_{\rm d}^{0.6}$ (combination describing the degeneracy line) as a function of redshift $z$. There is no evidence for evolution of the dust properties between $z=0$ and 2. The product increases between $z=2$ and 4.}
\caption{ \textit{Top panel:}  Average $\rm T_{\rm d} $ as a function of redshift $z$. \textit{Middle panel:} Average $\rm \beta_{\rm d} $ as a function of redshift $z$. There is a clear evolution between the low-z (z<1.5) , high-temperature and low-$\beta$ and the high-z, low-temperature and high-$\beta$ QSO populations. \textit{Bottom panel:} Average dust $\rm T_{\rm d} \times \rm \beta_{\rm d}^{0.6}$ (combination describing the degeneracy line) as a function of redshift $z$. There is no evidence for evolution of the dust properties between $z=0$ and 1.5. The product increases between $z=1.5$ and 4.}
\label{fig:dust_temp_all}
\end{figure}

We minimize the $\chi^2$ for each bin individually and plot the value for the combination $T_{\rm d} \times \beta_{\rm d}^{0.6}$ in Fig.~\ref{fig:dust_temp_all}. This quantity follows the degeneracy line for the dust parameters. There is no evidence for significant evolution across redshift between $z=0$ and 2.  For $z>2$, $T_{\rm d} \times \beta_{\rm d}^{0.6}$ increases with redshift. If we suppose that $\beta_{\rm d}$ remains constant, this implies that the dust was hotter at high redshift and cooled until $z=2$ before stabilizing at the current values.

\begin{figure}
\centering
\includegraphics[width=.5\textwidth]{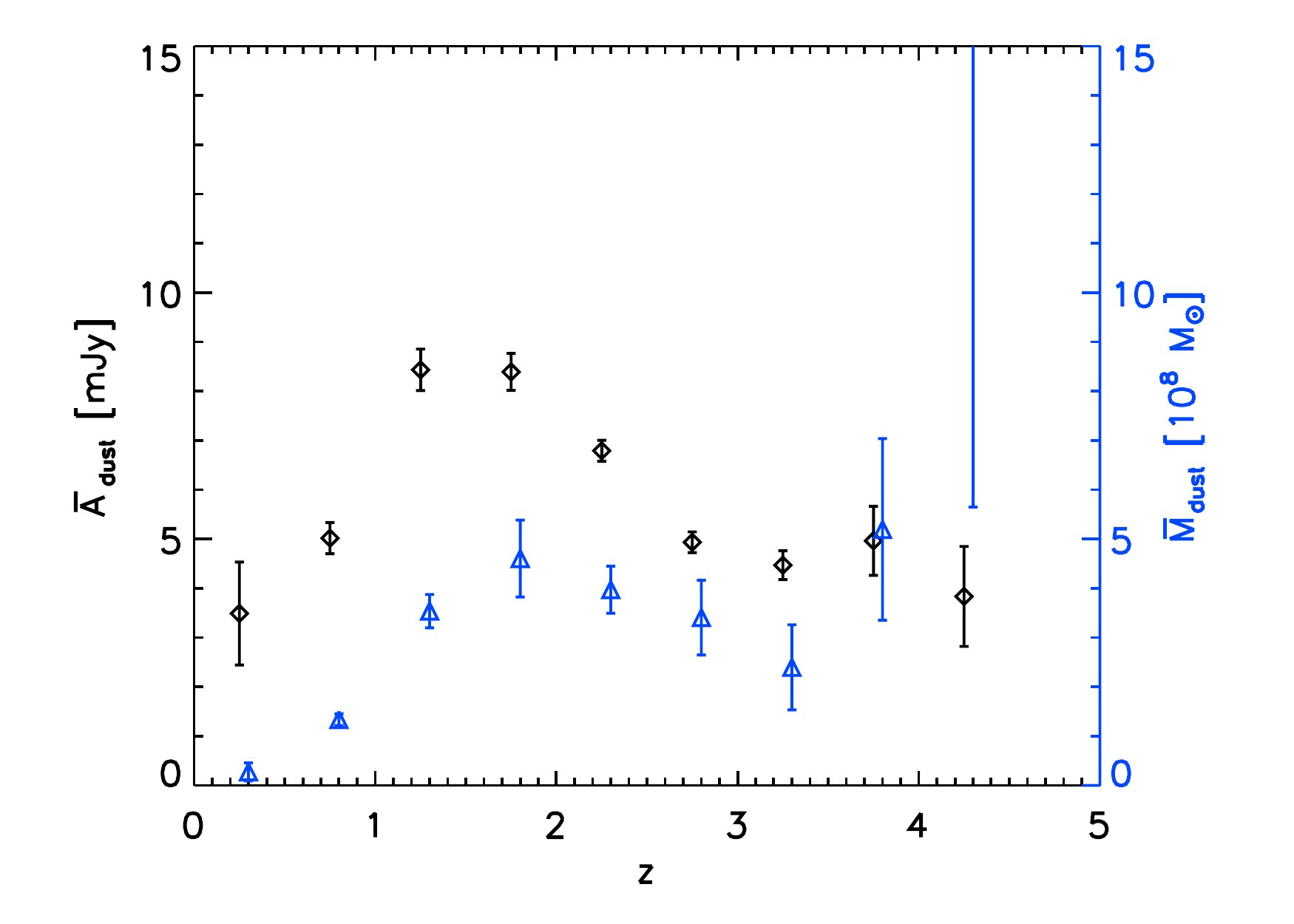}
\caption{Average dust flux density after marginalization over $\rm T_{\rm d}$ and $\rm \beta_{\rm d}$ (black diamonds) as a function of redshift $z$. The corresponding dust mass is shown as blue triangles (Eq.~\ref{eq:mdust}). The extraction is achieved using the dust-only filter.}
\label{fig:dust_mass_all}
\end{figure}

The dust flux density extracted using the dust-only MMF is shown in Fig.~\ref{fig:dust_mass_all} as the black diamonds. It varies with redshift, increasing from $z=0$ to $z=2$ ($\sim 8 \, {\rm mJy}$ within $R_{500}$ at 857~GHz) and decreasing with redshift for $z>2$. The corresponding dust mass (Eq.~\ref{eq:mdust}) is shown by the blue diamonds and follows the same trend as the flux density, reaching a maximum of $\sim 5 \times 10^8 M_\odot$ at $z \sim 2$.  This is of the same order as dust masses determined from high signal-to-noise observations of individual QSOs~\citep[e.g.,][]{2006ApJ...642..694B, wang2007}.

The IR luminosity, or equivalently the dust mass (Eq.~\ref{eq:mdust}), is a tracer of star formation, and it is remarkable that the dust mass evolution in Fig.~\ref{fig:dust_mass_all} follows that of the general star formation rate with a peak around $z\sim 2$ \citep[See, e.g., Fig. 9 in][]{madau2014}.  The accretion rate onto AGN SMBHs follows the same trend \citep{hopkins2007,shankar2009,aird2010,delvecchio2014}.  This is strong evidence that star formation in QSO environments is typical of the general galaxy population, and also that it is linked to the QSO central engines.  Our result clearly shows this trend for a large, representative sample of QSOs, and is consistent with the study by \citet{wang2015} of the correlation between QSOs and Herschel measurements of the cosmic infrared background.

Table~\ref{tab:DR12_data} also contains results for the dust+tSZ, dust+synch, and the dust+tSZ+synch filters on the full QSO population ($0.1<z<6.5$).  All three of the multiple component filters increase the $\chi^2$ with respect to the dust-only filter.

\begin{table*}
		\tiny
		\centering
		\caption{Average properties of different QSO samples in \Planck\ data, marginalized over $T_d$ and $\beta_d$ (see Sec.~\ref{sec:Marginalization on the (T,beta) plane}), except for line 1 and 6 marginalized over $T_d$ only, $\beta_d$ being fixed to 1.6 as in \cite{2006ApJ...642..694B}. QSO with at least one radio counterpart at 1.4GHz are flagged with 'yes' in the FIRST column. QSO with no identified counterpart are flagged with 'no'. For each sample, the filter leading to the smallest $\chi^2$ is highlighted in gray.  The different sample sizes are as follows: full sample ($0.1<z<5$), 291 165; restricted redshift ($2.5<z<4$), 90 576 of which 3354 have FIRST counterparts and 77 665 do not, and the remaining 9557 fall outside the FIRST coverage.
%		The full sample contains 291,232 QSOs spanning the redshift range $0.1<z<6.5$. Restricting the sample to $2.5<z<4$ provides 90,576 QSOs, with 3,354 FIRST (FIRST\_MATCH=1) and 77,665 non-FIRST (FIRST\_MATCH=0) objects. The remaining 90,576-3,354-77,665=9,557 QSOs are outside the FIRST survey coverage.
}
			\label{tab:DR12_data}
	\begin{tabular}{ccccccccccc}
		\hline
		\hline
		$z_{min}$ & $z_{max}$ & FIRST & Filter & $\chi^{2}$/5 &$T_{\rm d}$  & $\beta_{\rm d}$  &  $\overline{M_{\rm dust}} $ & $\overline{Y_{\rm 500}} $& $\overline{M_{\rm 500}} $&$\overline{L_{\rm synch}}$  \\
		  &  & & & &(K)  &  &($10^{8} \, M_\odot$) &$(10^{-6} \, arcmin^2)$ & $(10^{13} \, M_\odot)$& ($10^{-3} \, L_\odot {\rm Hz^{-1}}$)   \\
		\hline
		0.1 & 5 & - & dust & 16.3 & $28.2 \pm 0.5$ & $1.6$ (fixed) & $1.70 \pm 0.08$  & - & - &- \\
\rowcolor{lightgray}		0.1 & 5 & - & dust & 5.2 & $19.1 \pm 0.8$ & $2.71 \pm 0.13 $& $0.84 \pm 0.07$  & - & - &- \\
		0.1 & 5 & - & dust+tSZ & 9.1 & $18.6 \pm 0.8$ & $2.80 \pm 0.13$ & $0.76 \pm 0.07$  & $17.02\pm 0.82$  & $3.32 \pm 0.09$ &- \\
		0.1 & 5 & - & dust+synch & 6.7 & $20.0 \pm 1.0$ & $2.52 \pm 0.14$ & $0.98 \pm 0.09$  & - & - &$0.34 \pm 0.13$  \\
		0.1 & 5 & - & dust+tSZ+synch & 16.1 & $21.3 \pm 1.1$ & $2.24 \pm 0.15$ & $1.21 \pm 0.12$  & $17.15\pm 0.82$ &$3.34 \pm 0.09$ &$0.66 \pm 0.13$  \\
		\hline
		2.5 & 4 & - & dust & 12.4 & $28.7 \pm 0.5$ & $1.6$ (fixed) & $7.85 \pm 0.42$  & - & - &- \\
		2.5 & 4 & - & dust & 4.2 & $21.9 \pm 0.8$ & $2.63 \pm 0.14$ & $2.82 \pm 0.46$  & - & - &- \\
\rowcolor{lightgray}		2.5 & 4 & - & dust+tSZ & 2.3 & $22.5 \pm 0.9$ & $2.59 \pm 0.18$ & $2.60 \pm 0.53$  & $8.62 \pm 1.36$  & $2.26 \pm 0.20$ &- \\
		2.5 & 4 & - & dust+synch & 3.8 & $23.8 \pm 0.9$ & $2.25 \pm 0.16$ & $4.46 \pm 0.82$  & - & - &$-26.41 \pm 5.78$  \\
		2.5 & 4 & - & dust+tSZ+synch & 3.4 & $23.5 \pm 1.0$ & $2.34 \pm 0.22$ & $3.75 \pm 1.05$  & $5.92 \pm 2.25$ &$1.63 \pm 0.71$ &$-11.37 \pm 8.66$  \\		
		\hline
		2.5 & 4 & yes & dust & 4.9 & $32.5 \pm 5.1$ & $1.24 \pm 0.44$ & $16.31 \pm  6.40$  & - & - &- \\
		2.5 & 4 & yes & dust+tSZ & 2.0& $23.3 \pm 3.2$ & $2.01 \pm 0.46$  & $14.32 \pm 6.24$  & $ -40.14\pm 7.57$  & - &- \\
		2.5 & 4 & yes & dust+synch & 0.4 & $18.8 \pm 2.8$ & $3.72 \pm 0.67$  & $1.11 \pm 1.14$  & - & - &$236.91 \pm 30.47$  \\	
\rowcolor{lightgray}		2.5 & 4 & yes & dust+tSZ+synch & 0.3 & $15.7 \pm 3.3$ & $5.33 \pm 1.43$  & $0.38 \pm 0.68$  & $15.77 \pm 10.44$ &$2.23 \pm 1.88$ &$278.76 \pm 40.69$  \\		
		\hline
		2.5 & 4 & no & dust & 4.4 & $21.8 \pm 0.9$ & $2.69 \pm 0.19$ & $2.57 \pm  0.54$  & - & - &- \\
\rowcolor{lightgray}		2.5 & 4 & no & dust+tSZ & 1.5 & $22.4 \pm 1.2$ & $2.68 \pm 0.23$ & $2.18 \pm 0.56$  & $10.83 \pm 1.46$  & $2.58 \pm 0.20$ &- \\
		2.5 & 4 & no & dust+synch & 2.7& $24.0 \pm 1.0$ & $2.22 \pm 0.18$  & $4.47 \pm 0.89$  & - & - &$-34.17 \pm 6.23$  \\
		2.5 & 4 & no & dust+tSZ+synch & 2.3& $23.6 \pm 1.1$ & $2.34 \pm 0.23$  & $3.65 \pm 1.09$  & $6.82\pm 2.41$ &$1.83 \pm 0.67$ &$-16.86 \pm 9.24$  \\		
		2.5 & 4 & no & dust($\sigma_{\rm T}$=2 K)+tSZ & 2.1 & $20.9 \pm 1.1$& $2.81 \pm 0.22$ & $2.03 \pm 0.45$  & $10.55\pm 1.43$  & $2.54 \pm 0.20$ &- \\
		2.5 & 4 & no & dust($\sigma_{\rm T}$=5 K)+tSZ & 1.9 & $10.7 \pm 1.5$ & $3.61 \pm 0.23$  & $3.08 \pm 0.33$  & $9.37\pm 1.49$  & $2.37 \pm 0.21$ &- \\
		2.5 & 4 & no & dust($\sigma_{\rm T}$=10 K)+tSZ & 11.7 & $6.6 \pm 0.9$ & $2.31 \pm 0.02$ & $13.07 \pm 1.12$  & $9.73\pm 1.47$  & $2.42 \pm 0.21$ &- \\
		\hline
	\end{tabular}
\end{table*}

\subsection{Synchrotron emission}
\label{sec:synch}
\begin{figure}
\centering
\includegraphics[width=.5\textwidth]{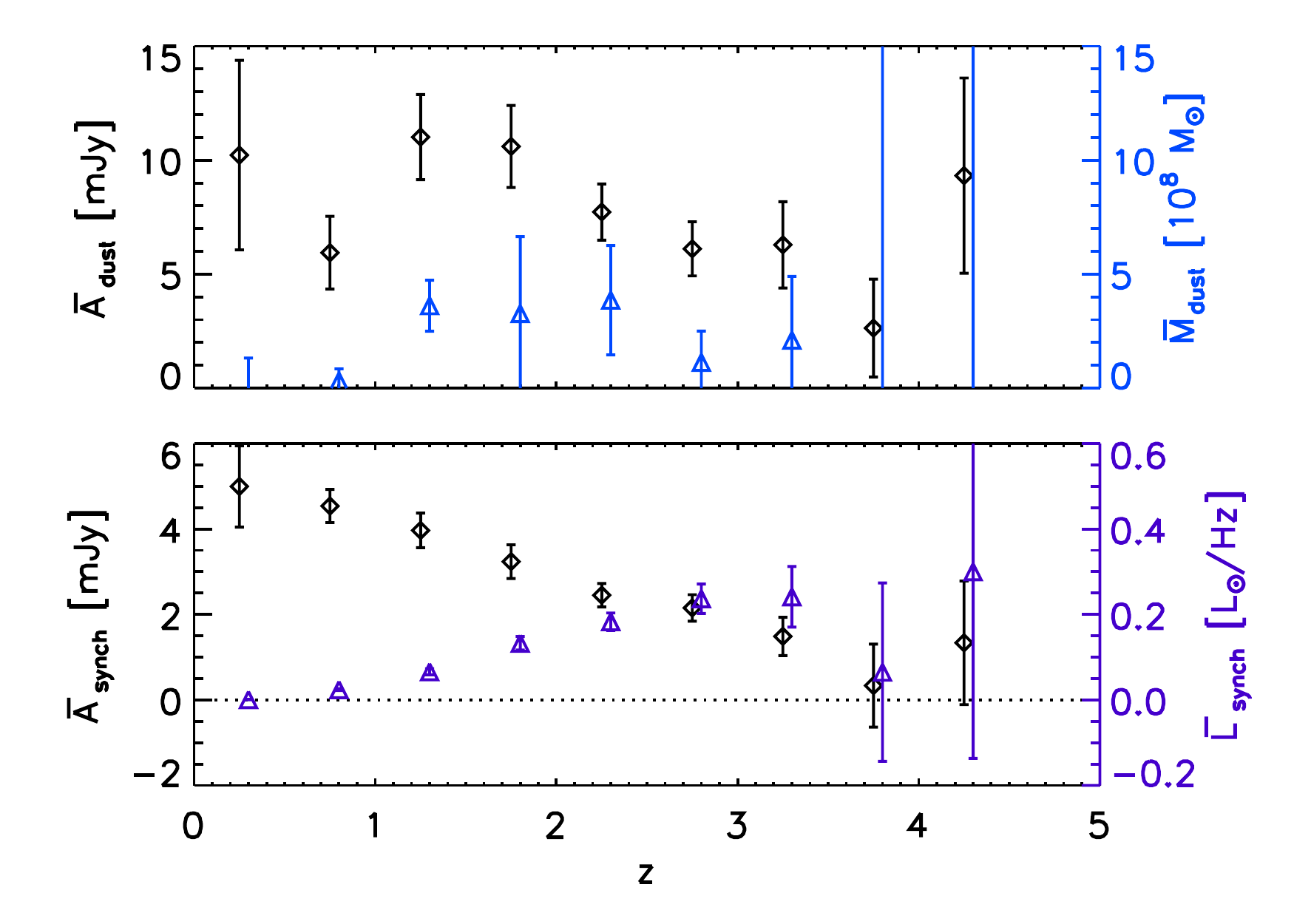}
\caption{\textit{Top panel:} Average dust flux density (black diamonds) for QSOs {\em with a radio counterpart in the FIRST catalogue}. The signal is extracted with the dust+synch filter and marginalized over $\rm T_{\rm d}$ and $\rm \beta_{\rm d}$. The corresponding dust mass is shown as the blue triangles.  \textit{Bottom panel:} Average synchrotron flux density at 100\,GHz observed frequency (black diamonds) and the corresponding monochromatic luminosity (magenta triangles).}
\label{fig:synch_z_radio}
\end{figure}

We now select only QSOs with at least one FIRST\footnote{\url{http://sundog.stsci.edu}} radio source counterpart at 1.4~GHz using the FIRST\_MATCH keyword in the DR12 catalogue (FIRST\_MATCH=1). This provides a sub-catalogue of 9983 QSOs.  Applying the dust+synch filter to this sub-catalogue using the same redshift binning, we plot the average dust and synchrotron flux densities, ${\bar A}_{\rm dust}$ and ${\bar A}_{\rm synch}$, in Fig.~\ref{fig:synch_z_radio}.  Dust emission from the sub-catalogue is compatible with that of the full catalogue, although with larger uncertainties due to the smaller sample (blue diamonds). 

This population of radio-loud QSOs exhibits strong synchrotron emission, as shown in the lower panel.  The monochromatic synchrotron luminosity (green diamonds) increases from 0 to reach 0.2 ${\rm L_\odot}\,{\rm Hz}^{-1}$ at the higher redshifts; We note that this is the luminosity at the rest-frame frequency $100$\,GHz, assuming our power-law in Eq.~({eq:lsynch}).

\begin{figure}
\centering
\includegraphics[width=.5\textwidth]{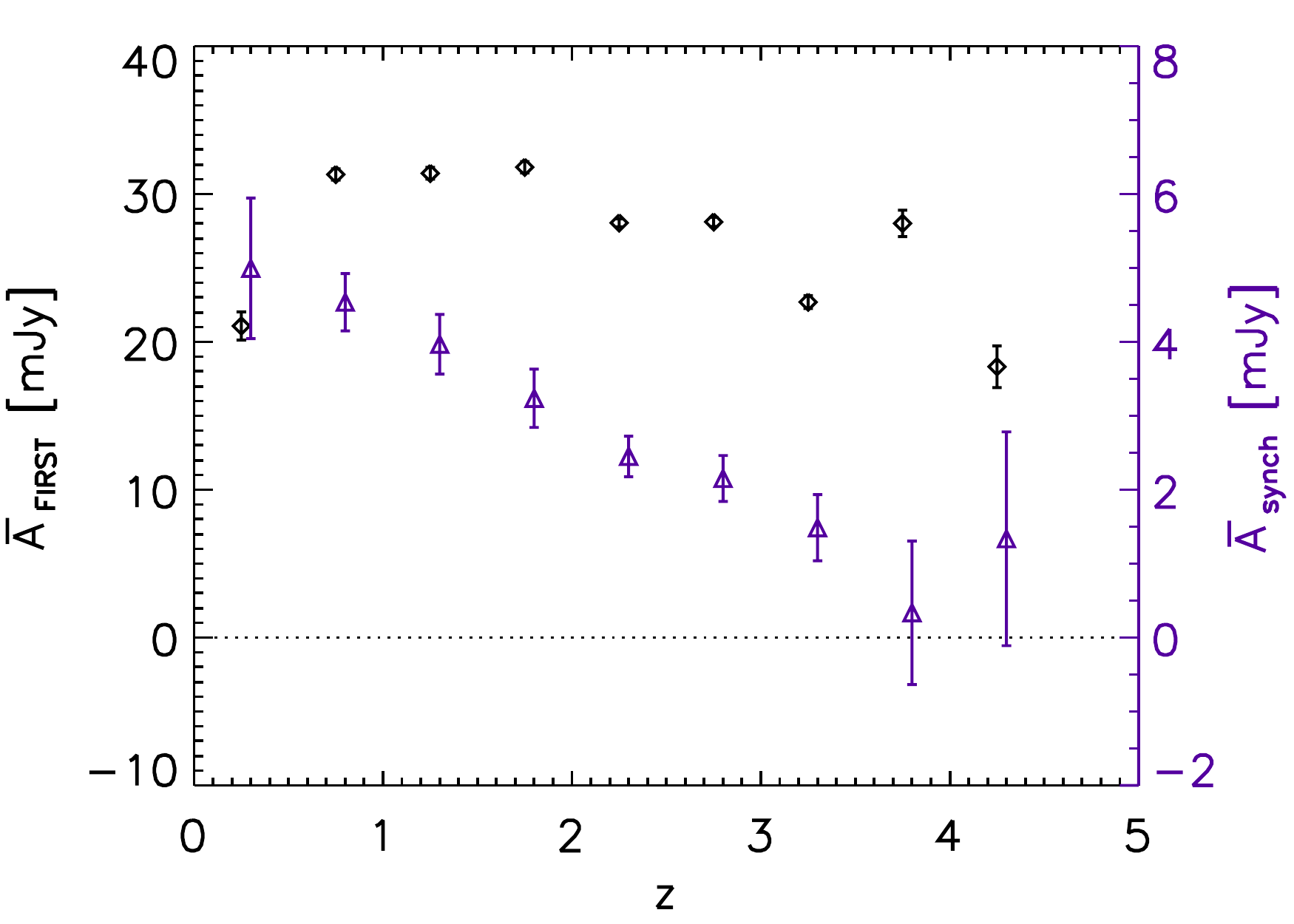}
\caption{Average FIRST flux density at 1.4\,GHz observed frequency compared to the average synchrotron flux density at 100\,GHz observed frequency as a function of redshift $z$. The magenta triangles correspond to the black diamonds in the bottom panel of Fig.~\ref{fig:synch_z_radio}.}
\label{fig:first_and_synch_z_radio}
\end{figure}

Fig.~\ref{fig:first_and_synch_z_radio} compares the synchrotron emission determined from \Planck\ (expressed at 100\,GHz) to the FIRST measurements at 1.4~GHz.  For this comparison, we bin-average the FIRST signal with the same MMF weights used to extract the synchrotron signal.  The emission seen by \Planck\ steadily decreases with redshift relative to the signal measured by FIRST, demonstrating a variation in synchrotron index with redshift.  
%The signal is still clearly present at $z=3$, where the 100\,GHz channel to which our synchrotron signal is normalized corresponds to a rest-frame frequency of 400\,GHz. 

\begin{figure}
\centering
\includegraphics[width=.5\textwidth]{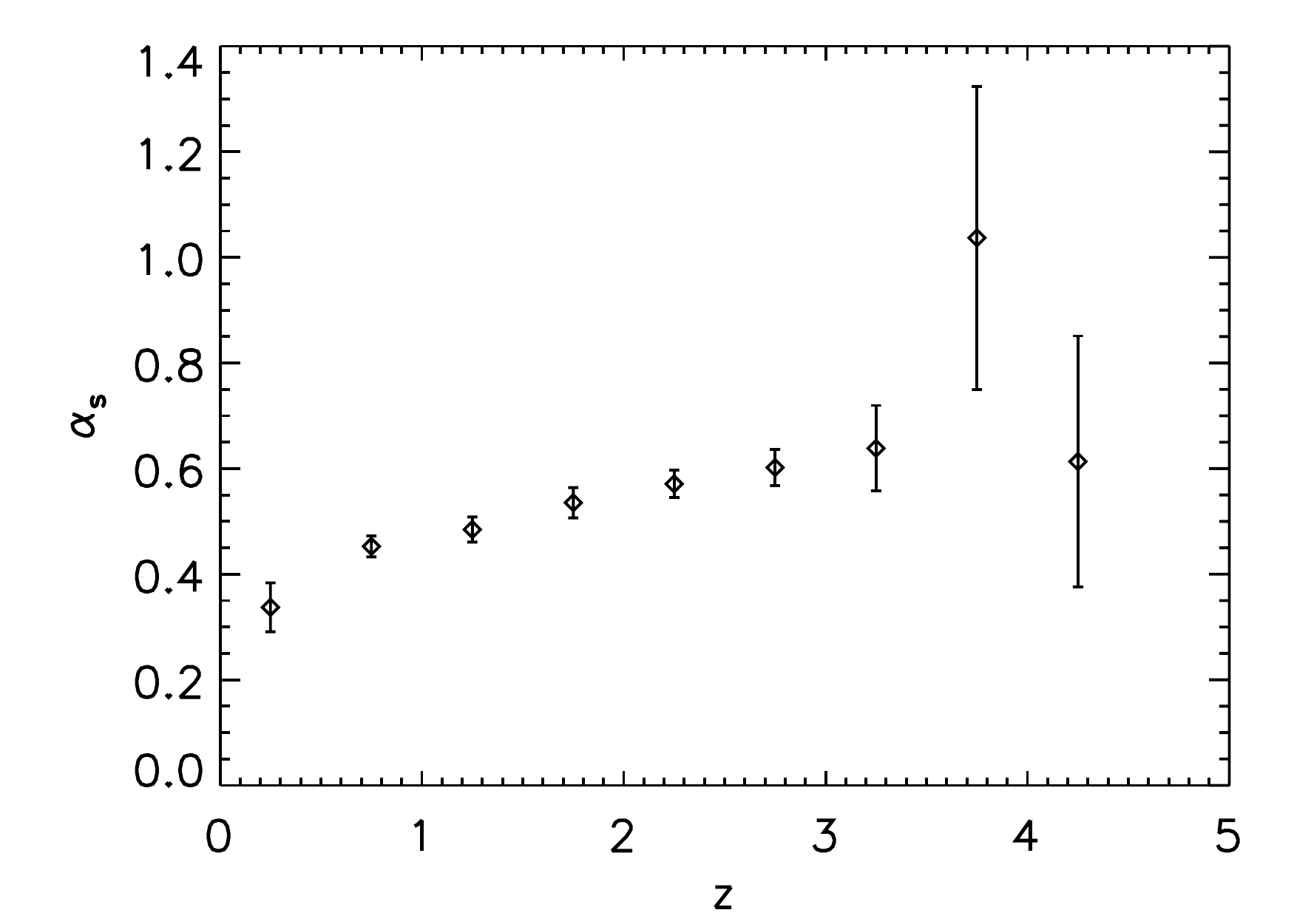}
\caption{Effective spectral index of power law in frequency between 1.4\,GHz (FIRST) and 100\,GHz (\Planck) observation frequencies.}
\label{fig:synch_alpha}
\end{figure}

We show the effective spectral index (for an assumed power law) between FIRST and \Planck\ in Fig.~\ref{fig:synch_alpha}.  The spectrum is quite flat at low redshift with values for the spectral index that are much smaller than our adopted value of $0.7$.  However, these are values describing the emission over a very large frequency range, while our adopted value is assumed to describe the emission around 100\,GHz.  

We also see that the effective spectral index steepens with redshift.  This could be due to evolution intrinsic to the sources, or more simply the effect of curvature in the synchrotron spectrum.  It is important to note in this light that although our synchrotron luminosity measurements  (Eq.~\ref{eq:lsynch}) are referenced to 100\,GHz rest-frame, assuming our adopted power law, the effective index shown in Fig.~\ref{fig:synch_alpha} is taken between the observed 1.4 and 100\,GHz bands.  In other words, it is the effective index between rest-frame frequencies of $1.4(1+z)$ and $100(1+z)$\,GHz.  The apparent evolution with redshift in the figure could therefore be steepening in the synchrotron emission with frequency that would be expected given the greater energy loss of the higher energy electrons contributing to the signal as we move up in rest-frame frequency.

%\jbm{Should we mention here the behavior as a function of g ?}\\
%\lv{There is a strong decrement of the dust flux and the synchrotron flux when g is increasing(when the optical QSO flux in band g is decreasing) but the associated masses don't change much across g. There is also no evolution of the halo mass. So we can mention at the end of part 5.1 that there is no significant physical evolution of our sources across the magnitude g.}

\subsection{Hot halo gas}
\label{sec:hotgas}

\begin{figure}
\centering
\includegraphics[width=.5\textwidth]{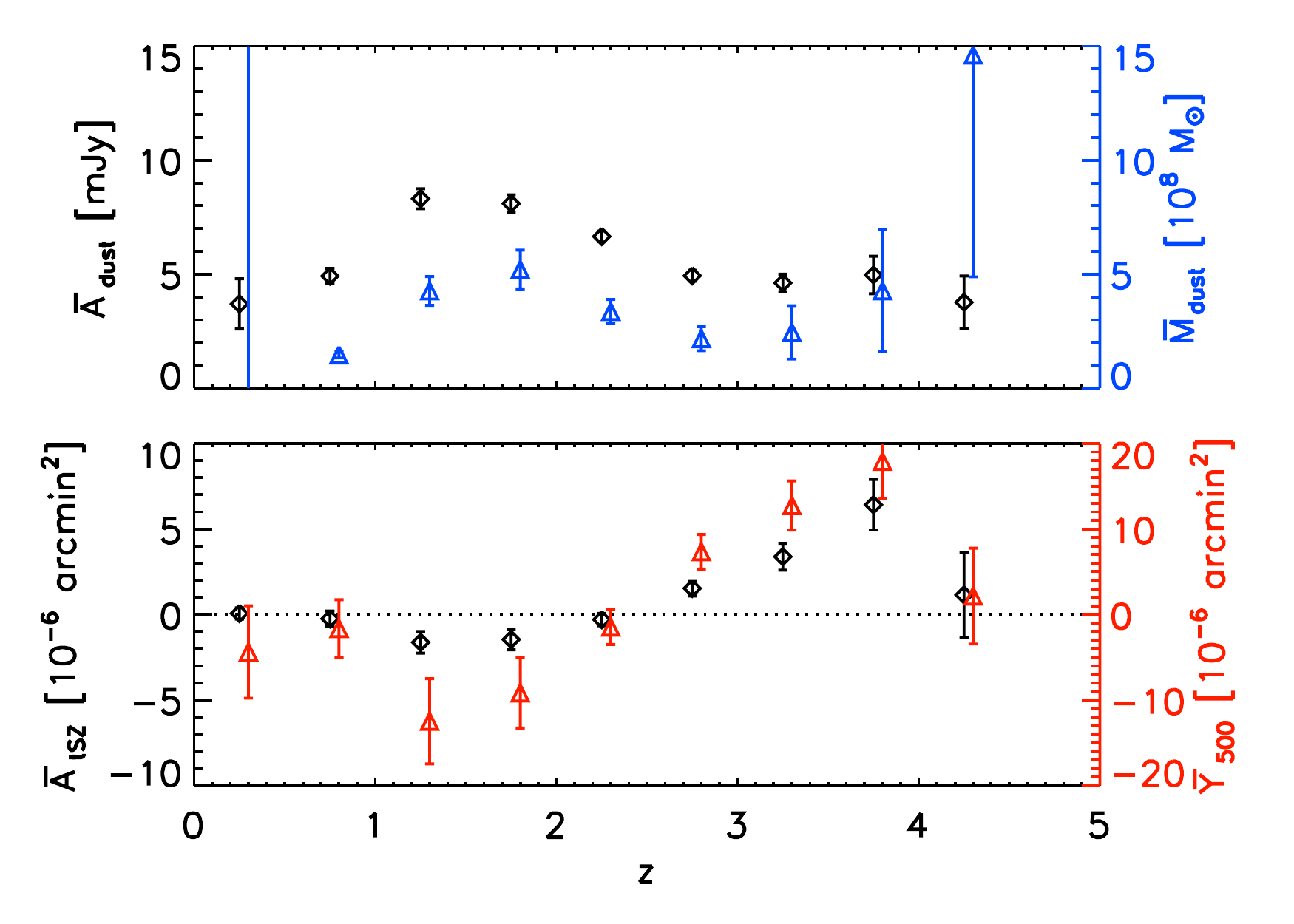}
\caption{\textit{Top panel:} Average dust flux density (black diamonds) of the radio-quite QSO subsample, i.e.,  without a radio counterpart in the FIRST catalogue. The signal is obtained with the dust+tSZ filter and is marginalized over $\rm T_{\rm d}$ and $\rm \beta_{\rm d}$. The corresponding dust mass is shown by the blue triangles. \textit{Bottom panel:} Average Compton parameter (black diamonds) and corresponding integrated Compton parameter (red triangles) as a function of redshift.}
\label{fig:sz_z_notradio}
\end{figure}

The strong synchrotron emission of QSOs with FIRST counterparts complicates extraction of the tSZ signal from the hot gas in the host halo. We therefore construct a sub-sample without FIRST counterparts (FIRST\_MATCH=0) of 252 888 QSOs and apply the dust+tSZ filter.  Results are shown in Fig.~\ref{fig:sz_z_notradio}, and summarized in the bottom segment of Tab.~\ref{tab:DR12_data} for this and other filters.

Applied to the same redshift range, the other filters yield larger $\chi^2$ values.  In particular, the dust+synch and dust+tSZ+synch filters show the same behavior as on simulations: they convert part of the tSZ signal into negative synchrotron luminosity.  The dust($\sigma_{T}$ =x K)+tSZ filters presented in Sect.~\ref{sec:syst} also increase the $\chi^2$, favoring a low scatter in the dust temperature. 

The filter does not detect the tSZ effect at $z<2.5$.  The tSZ signal at $z<1.5$ is slightly negative at $\sim 2\,\sigma$, pointing towards a possible contamination of the extracted signal at low $z$ by a weak synchrotron emission. The filter does show a clear signal rising with redshift over the range $2.5<z<4$.  The global significance of the signal in the latter redshift range is $7.4\sigma$ when expressed in terms of the population mean, $\overline{Y_{500}}$.  
%%
%Using the scaling relation Eq.~(\ref{eq:mhalo}), we derive an effective SZ-based halo mass for this redshift range of $\overline{M_{500}} \equiv (\estMfive^\alpha)^{1/\alpha} = (2.47 \pm 0.19) \times 10^{13} M_\odot = (1.73\pm 0.13) \times 10^{13} h^{-1} M_\odot$ ($h=0.7$).  %%We discuss this result further below.

The dust+tSZ+synch filter applied on the radio-loud sub-catalogue, i.e., with at least one FIRST source (third segment of Tab.~\ref{tab:DR12_data}), provides a slightly better fit ($\chi^2/5=0.3$) than the dust+synch filter ($\chi^2/5=0.4$) over the $2.5<z<4$ range.  The recovered SZ signal remains compatible with the value derived from the radio-quiet sub-catalogue, but the uncertainties are large with the signal appearing at only $1.5\sigma$.  

When applying the tSZ-only filter to the 2.5<z<4 population, we find SZ-based masses that are twice as large as the mass provided by the dust+tSZ filter; the fit, however, is quite bad at $\chi^2/7>100$ (We note that there are seven degrees-of-freedom in this case because the dust temperature and emissivity do not come into play).  This simply reflects the fact that a tSZ-only signal is a poor description of the spectrum of the source. Details of these results are given in Table~\ref{tab:DR12_data_SZ_only}.
%This confirms that dust is an strong contaminant for SZ measurements that must be properly accounted for.  The dust-tSZ filter achieves a good separation of the dust and tSZ signals, as attested its acceptable $\chi^2$ value.  

We stacked the ROSAT All-Sky Survey maps\footnote{\url{http://cade.irap.omp.eu/dokuwiki/doku.php?id=rass}} of the radio-quiet QSO sub-sample to find a significant signal, corresponding to a population mean luminosity of $L_{500}=(3.30 \pm 0.30) \times 10^{44} {\rm erg/s}$ in the [0.1-2.4] keV band, the error taken as the standard deviation of the stacked maps.  \citet{pratt2009} established a power-law scaling relation between X-ray luminosity and halo mass based on analysis of a representative sample of X-ray galaxy clusters employing Bivariate Correlated Errors and intrinsic Scatter estimators with a minimization of the residuals in X-ray luminosity (BCES (YlX)).  If we adopt their Malmquist-bias corrected relation between $L_{500}$, in this energy band, and mass (line 6 of their Table B.2), we infer a mean halo mass of $M_{500} = 5.87 \times 10^{13} M_\odot$.  This is 2.3 times higher than our SZ-based mass; alternatively, it implies an expected $\tilde{Y}_{500}= (51.4 \pm 4.6) \times 10^{-6} {\rm arcmin}^2$, or 4.7 times larger than our measurement. In other words, the tSZ and X-ray signals do not follow the empirical relation for thermal halo gas emission seen at low redshift \citep{2011A&A...536A..11P}.  This is not a surprise given that we expect a non-negligible contribution to the X-ray emission from the central engine of the QSO itself.
%We discuss the implications of our tSZ and X-ray measurements below.  

%The X-ray signal is consequently not composed by the intra-cluster gas emission only but also by an additional component, possibly the emission of the AGN.

%%
%This value is one order of magnitude above the halo mass of $10^{12} h^{-1} M_\odot$ given by the QSO clustering of the
%DR10 (\jbm{? TO BE CHECKED} \lv{QSO from BOSS with the spectrometry done before January 1, 2012. Non-standard(?) catalogue, post-DR7, close to DR9}
%BOSS sample in \cite{2012MNRAS.424..933W} at $z= 2.4$. But our sample is located in significantly higher redshift range ($2.5<z<4$). \cite{2007AJ....133.2222S} found that the QSO host haloes have a minimal mass $(2-3) \times 10^{12} h^{-1} M_\odot$ for $z \in [2.9,3.5]$ and $(4-6) \times 10^{12} h^{-1}  M_\odot$  for $z>3.5$. More recently, \cite{richardson2012} found QSO halo mass $ 14.1_{-6.9}^{+5.8} \times 10^{12} h^{-1} M_\odot$ at $z \sim 3.2$, a result which is both in agreement with \cite{2007AJ....133.2222S} and our derived mass.
%%

%For completeness' sake, we calculated the correlation between the derived tSZ halo mass and the dust mass for the dust+tSZ filter. The two quantities are slightly anti-correlated:
%\begin{equation}
%{M_{500}^{1.78} \over 5.01 \times (10^{13} M_\odot)^{1.78} } = \left ( M_{\rm dust} \over 2.15 \times 10^8 M_\odot \right )^{-0.27}
%\end{equation}

%\begin{figure}
%\centering
%\includegraphics[width=.5\textwidth]{./Figs/Plot_param_z_param2_radio0_Planck_DR12_ts_medium_filtered_maps.pdf}
%\includegraphics[width=.5\textwidth]{./Figs/Plot_param_z_param2_radio1_Planck_DR12_ts_medium_filtered_maps.pdf}
%\includegraphics[width=.5\textwidth]{./Figs/Plot_param_z_param2_radio3_Planck_DR12_ts_medium_filtered_maps.pdf}
%\caption{$\chi^{2}$ distribution across redshift for the three filters. \textit{Top panel:} \textit{Middle panel:} \textit{Bottom panel:}}
%\label{fig:sz_chi2_z}
%\end{figure}

\begin{table*}
		\centering
		\caption{Average properties from a SZ filtering for several samples of QSO.}
		\label{tab:DR12_data_SZ_only}
	\begin{tabular}{ccccccccccc}
		\hline
		\hline
		$z_{min}$ & $z_{max}$ & FIRST & Filter & $\chi^{2}$/7 & $\langle \widehat{Y}_{\rm 500} \rangle$ \\
		&  & & & &$(10^{-6} \, arcmin^2)$   \\
		\hline
		0.1 & 5 & - &  tSZ  & 520.8 & $20.25 \pm 0.81$ \\
		\hline
		2.5 & 4 & - &  tSZ  & 141.2 & $29.94\pm 1.17$ \\
		2.5 & 4 & yes &  tSZ  & 17.1 & $7.46 \pm 5.84$ \\
		2.5 & 4 & no &  tSZ  & 119.7 & $30.53 \pm 1.25$ \\
		\hline
	\end{tabular}
\end{table*}

\subsection{Sensitivity to the assumed size $\theta_{\rm s}$}
\label{sec:syssize}

We fix $\theta_{\rm s}= 0.27$ arcmin throughout our analysis. The value corresponds to a halo mass of $M_{500}=10^{13} \, M_\odot$ at $z=2$.  To test the robustness of our results to this assumption, we fix alternatively the size to $\theta_{\rm s}$=0.12 and $\theta_{\rm s}$=0.57\,arcmin, corresponding to halos of mass $M_{500}=10^{12} \, M_{\odot}$ and $M_{500}=10^{14} \, M_{\odot}$, respectively, at z=2. Results are shown in Tab.~\ref{tab:syst_theta}. Changing the filtering scale slightly increases the $\chi^2$, and only weakly impacts the derived parameters ($<0.5 \sigma$), leaving our results essentially unchanged.

\begin{table*}
		\centering
		\caption{Average properties of the $2.5<z<4$ QSO sample without FIRST counterpart for three filtering sizes $\theta_{\rm s}$.}
		\label{tab:syst_theta}
	\begin{tabular}{ccccccccccc}
		\hline
		\hline
		$z_{min}$ & $z_{max}$ & FIRST & $\theta$& Filter & $\chi^{2}$/5 &$T_{\rm d}$  & $\beta_{\rm d}$ &  $\overline{M_{\rm dust}} $ & $\overline{Y_{\rm 500}} $ & $\overline{M_{\rm 500}} $ \\
		&  & &(arcmin)& & &(K)  &  &  ($10^{8} \, M_\odot$) &$(10^{-6} \, arcmin^2)$  &$(10^{13} \, M_\odot)$  \\
		\hline
		2.5 & 4 & no & 0.12 &  dust+tSZ  & 2.0 & $22.3 \pm 1.0$ & $2.68 \pm 0.21$ & $2.00 \pm 0.47$  & $10.89\pm 1.42$ &$2.59 \pm 0.19$ \\
		2.5 & 4 & no & 0.27 &  dust+tSZ  & 1.5 & $22.4 \pm 1.2$ & $2.68 \pm 0.23$ & $2.18 \pm 0.56$  & $10.83 \pm 1.46$ &$2.58 \pm 0.20$\\
		2.5 & 4 & no & 0.57 &  dust+tSZ  & 2.6 & $22.6 \pm 1.0$ & $2.67 \pm 0.20$ & $2.23 \pm 0.50$  & $11.15 \pm 1.47$ &$2.62 \pm 0.20$ \\
		\hline
	\end{tabular}
\end{table*}

%______________________________________________________________

\section{Discussion}
\label{sec:discussion}
Submillimeter observations are a valuable source of information on the physical conditions in QSO environments because they probe star formation in and around the hosts and energy injection into the surrounding medium by the SMBHs.  Due to atmospheric absorption, this waveband is however difficult to access from the ground, and this has limited studies to relatively small samples of several tens of objects \citep{omont2001, isaak2002, omont2013, priddey2003, 2006ApJ...642..694B, wang2007, omont2013}.  

The space-based platforms WMAP, \Planck\ and \Herschel\ now open this window for much more extensive studies.  Taking advantage of \Planck's all-sky and wide-frequency coverage, we have determined several mean characterstics of the QSO population through binning measurements of dust, synchrotron and tSZ signals from the very large BOSS QSO sample.

\subsection{Dust emission}
We find that the QSO spectrum is dominated by graybody dust emission with a temperature of $T_{\rm d} =  19.1 \pm 0.8 \, {\rm K}$ and an emissivity spectral index of $\beta_{\rm d} =  2.71 \pm 0.13$ when averaged over the entire sample spanning the full redshift range, $0.1<z<5$; these are the one-dimensional marginalized constraints for the dust-only filter that gives the best $\chi^2$ (see Tab.~\ref{tab:DR12_data}).  While there is no clear evidence that these quantities evolve with redshift at $z<2$, there is a rise at higher redshifts in the product $T_{\rm d}\beta^{0.6}_{\rm d}$, describing the degeneracy direction, at higher redshifts (Fig.~\ref{fig:dust_temp_all}).  

Curiously, the temperature is notably lower than found in ground-based observations of high-redshift quasars, which find temperatures of 40-50\,K, while our value is more typical of ordinary galaxies like the Milk Way.  Our dust emissivity index is significantly steeper than the values of 1-2 for normal Galactic dust, and should be compared to the value of 1.6 found in the ground-based surveys.  It is notably steeper than the limiting case of $\beta=2$ for ideal insulators and conductors far from a resonance.  These characteristics remain in the other sub-samples and component filters summarized in Tab.~\ref{tab:DR12_data}, apart the unusual behaviour of the radio-loud sub-sample with FIRST counterparts.  Fixing $\beta_{\rm d}=1.6$, we obtain a somewhat higher mean temperature, $T_{\rm dust}=28.2\pm 0.5$\,K, but at the cost of a much worse spectral fit according to the large increase in $\chi^2$ (see Tab.~\ref{tab:DR12_data}).  

We calculate dust mass using the simple prescription of Eq.~(\ref{eq:mdust}), which can also be viewed as a measure of far-IR luminosity.  An important result is our finding that   %The average dust mass over the full redshift range $0.1<z<6.4$ is $\overline{M_{\rm dust}}=(0.82 \pm 0.07)  \times 10^{8} M_\odot$, a little lower than the values $few\times10^{8}\,M_\odot$ from ground-based studies.  As a function of redshift, we find that the 
the dust mass, a tracer of local star formation, evolves with redshift in the same way as the global cosmic star formation rate, peaking at $z\sim 2$~\citep[See e.g. Fig. 9 in][]{madau2014}.  Star formation in QSO environments apparently does not distinguish itself from others.  It is also known, from measures of AGN luminosity functions, that the SMBH accretion rate follows the same evolutionary track \citep{hopkins2007,shankar2009,aird2010,delvecchio2014}.  

\subsection{Synchrotron emission}
A subsample of our QSOs are radio loud with counterparts at 1.4\,GHz in the FIRST catalogue.  We detect synchrotron emission from this subsample with an monochromatic luminosity, referenced to rest frame frequency $100\,{\rm GHz}$ (adopting our power-law of Eq.~\ref{eq:lsynch}), increasing with redshift between 
$z=0$ and $z=3$.  Comparing the FIRST and \Planck\ flux densities, we find an effective spectral index between the two corresponding rest-frame frequencies that increases with redshift from $\alpha_{\rm s}=0.3$ to $\alpha_{\rm s}=0.7$.  This steepening is likely the result of curvature in the spectrum as we observe progressively higher rest-frame frequencies.  
  
\subsection{tSZ and feedback} 

Restricting analysis to the radio-quite sub-sample, we detect the tSZ effect (with the dust+tSZ filter) at $S/N \sim 7.4$ over the redshift range $2.5<z<4$, with an integrated Compton-y parameter of  $\tilde{Y}_{500}= (10.83 \pm 1.46) \times 10^{-6} {\rm arcmin}^2$ (one-dimensional, marginalized over dust properties).  We do not, however, find evidence of tSZ signal at the lower redshifts.  At $z<1.5$, we obtain a mean value of $\tilde{Y}_{500}(z<1.5)=(-4.93\pm 2.48)\times 10^{-6}$\,arcmin$^2$, which we convert simply into an upper limit at two sigma above zero as
\begin{equation}
\label{eq:tSZlimit}
\tilde{Y}_{500}(z<1.5) < 4.97 \times 10^{-6}\, {\rm arcmin}^2 \quad (2\sigma).
\end{equation}
Similarly, we find an upper limit at $z<2.5$ of 
\begin{equation}
\label{eq:tSZlimit_higherz}
\tilde{Y}_{500}(z<2.5) < 2.93 \times 10^{-6}\, {\rm arcmin}^2 \quad (2\sigma),
\end{equation}
even stronger, as expected given the behaviour seen in Fig.~\ref{fig:sz_z_notradio}.

\citet{ruan2015} recently reported a detection of the tSZ signal by stacking the \citet{hill2014} tSZ map, constructed from the \Planck\ frequency maps, on 26 686 spectroscopic DR7 QSOs.  For their low redshift bin at $z<1.5$, they find $\tilde{Y}^{\rm c}_{500} = (74\pm 30)\times 10^{-6}$\,arcmin$^2$, quoted in terms of the Compton-$y$ parameter integrated along the line-of-sight cylinder.  The conversion factor between the cylindrical and spherical integrated Compton-$y$ parameters for our adopted profile is $\tilde{Y}_{500}^{\rm c} = 1.52\times \tilde{Y}_{500}$, which translates the upper limit of Eq.~(\ref{eq:tSZlimit}) to $\tilde{Y}_{500}^{\rm c} < 7.55\times 10^{-6}$\,arcmin$^2\,  (2\sigma)$, well below their detection level.  
 
%\jim{Almost all of their QSOs are at z<2.5; what's our upper limit at $z<2.5$?}
%\jbm{Our upper limit at $z<2.5$ is
%\begin{equation}
%\label{eq:tSZlimit_higherz}
%\tilde{Y}_{500}(z<2.5) < 2.93 \times 10^{-6}\, {\rm arcmin}^2 \quad (2\sigma).
%\end{equation}
%}
One possible explanation for the difference is dust contamination.  \citet{ruan2015} took particular pains to correct for any contamination by dust in the tSZ map.  Our filters directly measure the individual local signals from dust and tSZ, as well as any covariance between them, allowing us to separate and marginalize over the dust influence when quoting our tSZ signal strengths.  The importance of separating out the dust signal is easily appreciated from the results of applying the tSZ-only filter given in Tab.~\ref{tab:DR12_data_SZ_only}.  Contamination by dust in the tSZ-only filter can produce tSZ signals $\sim 3$ times larger.  

The tSZ signal directly measures the thermal energy stored in diffuse ionized gas in the QSO environment.  We expect a signal from the intra-halo gas at the virial temperature of the QSO host halo, and there may be additional heating powered by feedback from the SMBH.  \citet{chatt2010} and \citet{ruan2015} both interpreted their tSZ signals as evidence for feedback into the surrounding gas.  

To look for this, we estimate the tSZ signal expected from intra-halo gas using the scaling relation between halo mass and tSZ signal in Eq.~(\ref{eq:mhalo}).  We first establish a mean mass scale for the sample of 77 655 QSOs in $2.5<z<4$ over the 14 555 sq. deg. SDSS footprint.  Using the \citet{tinker2008} mass function, we find the lower mass bound above which there are as many predicted halos as observed QSOs in this redshift range and for this sky area.  We then calculate the mean mass by integrating over the mass function above this mass limit and over this redshift range.  This gives us an estimate of the maximum mean mass characterizing the sample; QSOs could be hosted by lower mass halos if the QSO selection function is not 100\% complete, as is most certainly the case.  

We find a mean halo mass of $M_{500, {\rm true}}=2.18\times 10^{13}\, M_\odot$, a value in reasonable agreement with \cite{richardson2012}, who found a QSO halo mass $14.1_{-6.9}^{+5.8} \times 10^{12} h^{-1} M_\odot$ at $z \sim 3.2$ for the SDSS sample.  Our characteristic mass corresponds to a predicted tSZ signal of $Y_{500}=8.47(1-b)^{1.78}\times 10^{-6}$\,arcmin$^2$, only one sigma below our measured signal of $Y_{500}=(10.02\pm 1.34)\times 10^{-6}$\,arcmin$^2$ if $b=0$.  This would not leave much room for additional heating by feedback.  With the higher mass bias values suggested by \Planck's cluster counts, e.g., $(1-b)=0.6$, we instead have $Y_{500}=3.4\times 10^{-6}$\,arcmin$^2$, implying that $\sim 60$\% of the gas's thermal energy is generated by feedback.

%Our characteristic mass corresponds to a predicted tSZ signal of $Y_{500}=8.74(1-b)^{1.78}\times 10^{-6}$\,arcmin$^2$, only one sigma and half below our measured signal of $\overline{Y_{\rm 500}} =(10.83\pm 1.46)\times 10^{-6}$\,arcmin$^2$ if $b=0$.  This would not leave much room for additional heating by feedback.  With the higher mass bias values suggested by \Planck's cluster counts, e.g., $1-b=0.6$, we instead have $Y_{500}=4.63\times 10^{-6}$\,arcmin$^2$, implying that $\sim60$\% of the gas's thermal energy could be supplied by feedback. Our mean halo mass of $\overline{M_{\rm 500}}=2.58 \pm 0.20 \times 10^{13}\, M_\odot$, derived from our measured $\overline{Y_{\rm 500}}$, is in reasonable agreement with \cite{richardson2012}, who found a QSO host halo mass of $14.1_{-6.9}^{+5.8} \times 10^{12} h^{-1} M_\odot$ at $z \sim 3.2$ for the SDSS sample. 

X-ray emission from the hot gas is also expected.  Stacking the RASS maps at the QSO position, we find a luminosity $L_{500}=(3.30 \pm 0.30) \times 10^{44} {\rm erg/s}$ in the [0.1-2.4]\,keV band.  The tSZ -- X-ray luminosity scaling relation established by \citet{pxcc2011}  says that this should correspond to an SZ signal of $\tilde{Y}_{500}= (51.4 \pm 4.6) \times 10^{-6} {\rm arcmin}^2$, much higher than our actual tSZ measurement, a conclusion that is independent of the mass bias.  

The cluster halo gas scaling relations appear therefore to be violated in the QSO environment.  This could be due to feedback from the QSO SMBH.  It is likely that X-ray emission from the QSO itself also affects our X-ray measurements.  Moreover, we are extrapolating the relation to redshifts much higher than those used to establish its validity.  The redshift evolution of these relations is in fact not well known, and we have adopted self-similar evolution as a reasonable assumption given the lack of empirical constraints.  Measurements of these relations on non-QSO systems at the same redshifts would provide a valuable comparison sample to judge if the QSO's activity is in fact responsible for violating the scaling relations. 

Finally, we note that interpretation of the measured tSZ signal is also clouded by possible contribution from gas in structures correlated with the QSO halos and falling with the relatively large effective beam of {\em Planck}, as suggested by \cite{2015ApJ...809L..32C}.  We leave careful modeling of this effect to future work. 

\section{Conclusion}
\label{sec:conclusion}
\Planck's coverage of the SDSS survey area gives us the opportunity for in-depth study of quasar environments through submillimeter observations of optically selected QSOs.  This spectral band provides valuable information on star formation and energy production by the QSO central engines.  Filtering \Planck's seven spectral bands from 70-857\,GHz with multi-component MMFs, we have measured dust, synchrotron and tSZ signals from the large BOSS QSO sample spanning redshifts out to $0.1<z<5$.  While measurements of individual QSOs are well below \Planck's sensitivity, the MMFs enable us to determine the population's mean properties as a function of redshift.

Dust emission dominates the average QSO spectrum, and its evolution with redshift suggests that star formation in QSO environments follows the general cosmic star formation rate with a peak at $z\sim 2$.  We did not detect a tSZ signal from QSOs at $z<2.5$, but did find a signal at higher redshifts.  The signal is clearly seen out to $z\sim 4$, the highest redshift to which the tSZ signal has been measured to date.   A non-negligible fraction of the gas's thermal energy could be supplied by QSO feedback, but interpretation is difficult because of modeling uncertainties associated with halo gas scaling relations.  Finally, we observe synchrotron emission from a radio-loud sub-sample at rest-frame frequencies $\nu>100$\,GHz and whose luminosity rises monotonically with redshift.  

These results are based on a large sample of nearly 300 000 BOSS spectroscopic QSOs falling outside of the \Planck\ mask.  Our findings present a fresh and representative view of the average properties characterizing the QSO population. 

\begin{acknowledgements}
Some of the results in this paper have been derived using the HEALPix \citep{2005ApJ...622..759G} package.
Part of this research was carried out at the Jet Propulsion Laboratory, California Institute of Technology, under a contract with the National Aeronautics and Space Administration.
We thank T. Marriage for helpful discussions.
\end{acknowledgements}

\bibliographystyle{aa}
\bibliography{gasqso,gasqso_jgb}

%-------------------------------------------------------------------

\end{document}